\documentclass[journal]{IEEEtran}

\usepackage{graphicx}
\usepackage{float}
\usepackage{dirtytalk}
\usepackage{amsmath}
\usepackage{algorithm} 
\usepackage{algpseudocode} 
\usepackage[dvipsnames]{xcolor}
\usepackage{array}
\usepackage{multirow}
\usepackage{footmisc}
\usepackage{hyperref}
\usepackage{cite}
\usepackage{stfloats} 

\ifCLASSINFOpdf

\else
  
\fi

\begin{document}

\title{On the Peak-to-Average Power Ratio of Vibration Signals: Analysis and Signal Companding for an Efficient Remote Vibration-Based Condition Monitoring}

\author{Sulaiman~Aburakhia,~\IEEEmembership{Student Member,~IEEE,}
      and~Abdallah~Shami,~\IEEEmembership{Senior~Memeber,~IEEE}%

\thanks{S. Aburakhia and A. Shami are with the Department of
Electrical and Computer Engineering, Western University, 
N6A 3K7, Canada (e-mail: saburakh@uwo.ca; abdallah.shami@uwo.ca).}}

\markboth{}%
{Shell \MakeLowercase{\textit{et al.}}: Bare Demo of IEEEtran.cls for IEEE Journals}

\maketitle

\begin{abstract}
Vibration-based condition monitoring (VBCM) is widely utilized in various applications due to its non-destructive nature. Recent advancements in sensor technology, the Internet of Things (IoT), and computing have enabled the facilitation of reliable distributed VBCM where sensor nodes are deployed at multiple locations and connected wirelessly to monitoring centers. However, sensor nodes are typically constrained by limited power resources, necessitating control over the peak-to-average power ratio (PAPR) of the generated vibration signals. Effective control of PAPR is crucial to prevent nonlinear distortion and reduce power consumption within the node. Additionally, avoiding nonlinear distortion in the vibration signal and preserving its waveform is essential to ensure the reliability of condition monitoring. This paper conducts an in-depth analysis of the PAPR of vibration signals in VBCM systems, evaluates the impact of nonlinear power amplification on the system performance, and proposes a lightweight autoencoder-based signal companding scheme to control the PAPR to improve power efficiency and mitigate the impact of nonlinear distortion. The proposed scheme employs a lightweight reconstruction autoencoder with a compression-based activation function in the source to compress the vibration signals and avoid increasing the average power of the compressed signal. In the destination, the proposed scheme uses a denoising-expansion autoencoder to expand the compressed signals while minimizing noise enhancement during the expansion process. The experimental results demonstrate the effectiveness of the proposed companding scheme in preventing nonlinear distortion, improving the efficiency of power amplification in the source, and restoring the PAPR characteristics in the destination while avoiding the undesired effect of noise expansion.

\end{abstract}

\begin{IEEEkeywords}
vibration-based condition monitoring (VBCM), sensors, power efficiency, peak-to-average power ratio (PAPR), signal companding
\end{IEEEkeywords}

\IEEEpeerreviewmaketitle

\section{Introduction}

\IEEEPARstart{V}{ibration}-based condition monitoring (VBCM) has been widely used in predictive maintenance (PdM) \cite{mech0}\cite{mech1} and structural health monitoring (SHM) \cite{shm0}\cite{shm1}. Furthermore, it is becoming increasingly popular, primarily due to its inherent advantages over alternative forms of condition monitoring. The main advantages of VBCM include \cite{hos20}\cite{aircraft}:
\begin{itemize}
    \item  Vibration sensors are non-intrusive and can be contactless, facilitating non-destructive condition monitoring.
    \item  Real-time acquisition of vibration signals can be conducted in situ, allowing for online local condition monitoring.
    \item Trending vibration analysis can be utilized to identify relevant conditions and conduct comparative analysis across diverse conditions or objects.
   \item  Vibration sensors are cost-effective and widely available, offering various specifications to suit a wide range of requirements.
   \item Vibration waveform responds instantly to changes in the monitored condition and, therefore, is suitable for continuous and intermittent monitoring applications. 
   \item Of paramount significance, signal processing techniques can be applied to vibration
signals to mitigate corrupting noise and extract weak condition indications from other masking signals.
\end{itemize}

The rapid evolution of sensor fabrication, coupled with advancements in the Internet of Things (IoT) and computing technologies, has enabled the facilitation of large-scale remote VBCM systems comprising distributed sensor nodes. These systems are being widely utilized across diverse domains, including civil applications, industry, wildlife monitoring, agriculture, transportation, and healthcare applications. \cite{app1, app2, app3, app4, app5, jdm23, kim22, pen23, mun23, ars23, don23, zha21, sin23, kar23, sha23}. In these systems, sensor nodes are typically battery-powered and, therefore, have limited power resources \cite{iot3, iot4, vbm1, vbm2}. Hence, efficient power utilization in key components of the node, such as signal acquisition, amplification, and transmission, is paramount to maintaining low power consumption. Aiming at reducing power consumption, current research efforts are mainly dedicated to developing power-efficient signal acquisition techniques. Specifically, compressive sensing (CS) \cite{cm0, cm01, cm1, cm2, cm3, cm4, cm5, cm6, cm7, cm8, cm9} is being adopted to acquire the signal in a compressed form by performing fewer measurements and collecting fewer samples. This contrasts traditional sampling techniques that rely on the Nyquist-Shannon sampling theorem and require more measurements and samples. The ultimate objective is to achieve more power-efficient signal acquisition with less power consumption. CS allows the reconstruction of the signal for a few acquired samples However, reconstruction algorithms involve using sparse optimization and relying on the acquired signal's sparsity to reconstruct it from these fewer samples. As a result, the practical use of CS is limited by the assumption of signal sparsity and the costly reconstruction process \cite{cm10} that involves time and power-consuming algorithms, making CS unsuitable for real-time condition monitoring \cite{cm11 }. \par

This paper tackles the problem of power efficiency from a signal waveform perspective. Specifically, the paper suggests reducing the power consumption in the sensor nodes by controlling the peak-to-average power ratio (PAPR) of the acquired vibration signal. The PAPR, which is the ratio of the peak power to the signal's average power, has a direct impact on the node's power consumption since it determines the required resolution for analog-digital conversions \cite{ad0} as well as the required linear range of the power amplification circuit \cite{pa},  which accounts for the major part of the total power consumption in many systems \cite{pa3, pe1}. To the best of our knowledge, this paper is the first work that addresses the issue of PAPR in vibration signals and tackles the related problem of nonlinear power amplification in VBCM systems. Specifically, the paper statistically investigates the PAPR characteristics of vibration signals, evaluates the impact of nonlinear power amplification on the system, and proposes a lightweight framework based on signal companding to reduce the PAPR and ensure linear power amplification of the signals. Companding\footnote[1]{The name COMPANDING is a composite of the words COMPressing and expANDING.} is a well-known technique in signal processing; it involves signal compression at the source and subsequent expansion at the destination. Signal companding has demonstrated its effectiveness in controlling the PAPR of multi-carrier communication signals. Nevertheless, conventional companding techniques encounter two significant limitations. Firstly, the compression mechanism increases the average power of the compressed signal. Secondly, the expansion operation amplifies the accumulated noise in the compressed signal. In order to effectively control the PAPR and address these limitations, the proposed framework adopts a two-fold approach. Firstly, it compresses the signal with a reconstruction autoencoder with a compression-based activation function. Secondly, it employs a denoising-expanding autoencoder to expand the compressed signal. This combined approach ensures efficient PAPR control while mitigating the aforementioned limitations. The paper makes the following key contributions:
\begin{itemize}

\item To the best of our knowledge, this paper is the first contribution to the VBCM literature that addresses the PAPR of generated vibration waveform, examines the impact of nonlinear power amplifications, and proposes controlling the PAPR to improve power efficiency and mitigate nonlinear distortion. 
\item Statistically analyzes the PAPR of vibration signals. Accordingly, a closed-form formula is derived to accurately model the statistical distribution of the PAPR of vibration Signals.
\item Introduces a framework based on signal companding to effectively reduce the PAPR of vibration signals and mitigate the impact of nonlinear power amplification.
\item At the sensor node (source), prior to the power amplification stage, the framework employs a lightweight reconstruction autoencoder that utilizes a compression-based activation function. The autoencoder function facilitates the simultaneous smoothing and compression of the vibration signal without causing an increase in the average power of the compressed signal.
\item  At the monitoring end (destination), the proposed framework utilizes a denoising-expansion autoencoder to effectively remove noise from the compressed signals prior to the expansion process to avoid enhancement (expansion) of the accumulated noise by the expansion operation.
\item  Comprehensively evaluates the performance of the proposed framework in the presence of nonlinear power amplification and additive white Gaussian noise (AWGN), employing a real-world vibration dataset.
\item Adapts the concepts of signal constellation diagrams and error vector magnitudes (EVM) as new metrics to evaluate power efficiency and quantify the nonlinear distortion resulting from the nonlinear power amplification.
\end{itemize}

The remainder of the paper is structured as follows: The next section provides background information and motivation for the problem. Section 3 presents the statistical analysis of the PAPR of vibration signals. Section 4 reviews signal companding techniques that have been proposed in the literature. Section 5 introduces the proposed autoencoder-based companding framework. Section 6 presents the model of nonlinear power amplification used in the experimentation. 
The experimental setup and performance evaluation metrics are introduced in Section 7, while Section 8 discusses the obtained results. The paper is finally concluded in Section 9.

\section{Background and Motivation}

\begin{figure*}[!htbp]
\centerline{\includegraphics[width=0.8\textwidth]{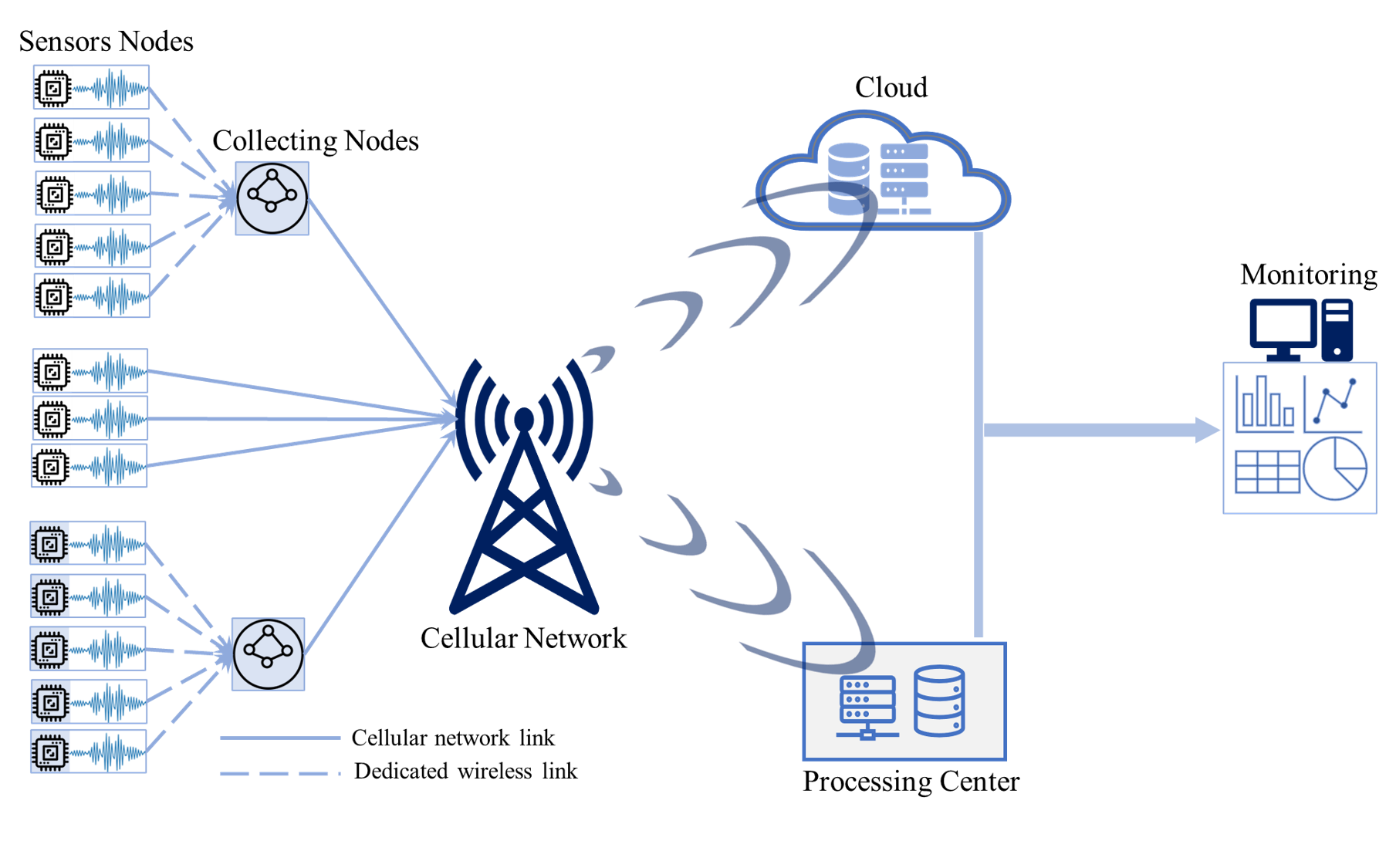}}
\caption{Overview of a typical remote VBCM system.}
\label{fig7.1.1}
\end{figure*}

Fig. \ref{fig7.1.1} shows a high-level architecture of a typical remote VBCM system. 
In these systems, sensor nodes are deployed across various locations, either embedded in objects, placed beneath surfaces, attached to mobile or airborne objects and connected to a cloud or a processing center. Typically, sensor nodes use wireless connectivity through available cellular networks or dedicated wireless links to collecting nodes in cases where cellular coverage is unavailable or unstable \cite{app4, app5, iot1, iot2, iot3, iot4, iot5}. Subsequently, the collection node transmits the accumulated signals to the cloud or the processing center via the cellular network. As mentioned earlier, these nodes are typically power-constrained, which makes efficient power utilization in key components of the node, such as signal accusation, amplification, and transmission, critical for low power consumption in these nodes. \par

\begin{figure}[!htbp]
\centerline{\includegraphics[width=0.5\textwidth]{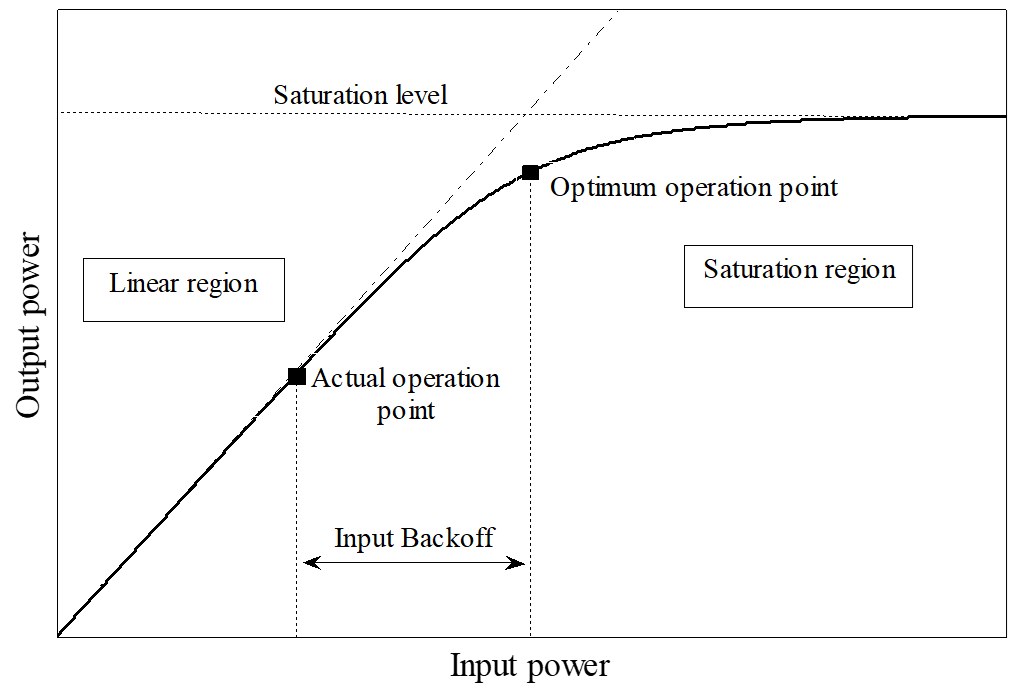}}
\caption{HPA range curve.}
\label{fig7.1.2}
\end{figure}

As stated earlier, the PAPR is crucial in determining how much power the system needs to operate effectively. A smaller PAPR value requires fewer bits and allows HPA to operate more efficiently, saving the battery in the system \cite{pe2}. To achieve maximum power efficiency, the HPA's operating point should be positioned as close as possible to HPA's saturation point \cite{pe3} as illustrated in Fig \ref{fig7.1.2}. 
When input signal peaks exceed this designated operating point, the HPA becomes prone to saturation, leading to power wastage, nonlinear amplitude distortion, and spectral spreading induced by abrupt fluctuations in the distorted amplitudes. To prevent these consequences, the HPA circuit must be designed to operate linearly over the PAPR range of the input signal, which tends to be a costly and inefficient solution \cite{pa4}. Alternatively, a significant input power backoff (IBO) from the HPA's operating point should be applied to restrict the HPA's input power level, ensuring that the entire signal falls within the HPA's linear region. While this approach mitigates nonlinear distortion, it significantly reduces power efficiency, as the HPA operates in a lower-power region. Therefore, it has a high cost in terms of energy efficiency, particularly in battery-power applications \cite{pe3} such as remotely deployed sensor nodes.

In VBCM systems, the amplitude of the acquired vibration waveform fluctuates according to the condition of the monitored object/process. Hence, the waveform is anticipated to exhibit a high PAPR due to these fluctuations, which can reach significant magnitudes depending on the monitored condition \cite{app4}\cite{app5}. 

Based on the preceding discussion, the remainder of the paper attempts to facilitate power-efficient remote VBCM by addressing the following key aspects:
\begin{enumerate}
    \item Analyse PAPR characteristics of acquired vibration signals,
    \item evaluate the impact of uncontrolled PAPR on the VBCM performance in the presence of nonlinear power amplification, 
    \item and, accordingly, propose appropriate remedy solutions to control the PAPR.
\end{enumerate}
\section{PAPR of Vibration Signals}

The PAPR is a widely employed metric for quantifying the power ratio between the peak and average values of a signal. For a given vibration signal $x(t)$, its PAPR can be expressed as:
\begin{equation}
PAPR_{x(t)}=\frac{Max\left | x(t)\right |^2}{E\left \{ \left | x(t)\right |^2 \right \}}
\label{papr}
\end{equation}
where  $E\left \{ \cdot  \right \}$ denotes the expectation operator. For the finite sampled signal $x(n)$, the PAPR is:
\begin{equation}
PAPR_{x(n)}=\frac{Max_{n\epsilon [0,N]}\left | x(n) \right|^2}{\frac{1}{N}\sum_{0}^{N-1}\left | x(n) \right|^2}
\end{equation}
where  $N$ is the number of samples in the vibration signal $x(n)$. The PAPR is usually expressed in dB:
\begin{equation}
PAPR \,\text{(dB)} =10\times log\left ( PAPR \right ) \text{dB}
\end{equation}
Crest Factor (CF) is another common signal parameter that quantifies a signal's peak amplitude to its root-mean-square (RMS) value. It equals to the square root of the PAPR. However, expressed in dB, the CF is equal to the PAPR since:
\begin{equation}
CF \,\text{(dB)} = 10\times log\left ( CF^2 \right )\text{dB}=10\times log\left ( PAPR \right ) \text{dB}
\end{equation}
A signal with constant power, such as a square wave, has a PAPR of 1 ($0$ dB). The PAPR of a sinusoidal wave equals $2$ or $3.01$ dB. Determining the PAPR of a random vibration depends on its instantaneous value, which is not predictable. Nevertheless, it is possible to describe the PAPR statistically. 
\begin{figure*}[!htbp]
\centerline{\includegraphics[width=1\textwidth]{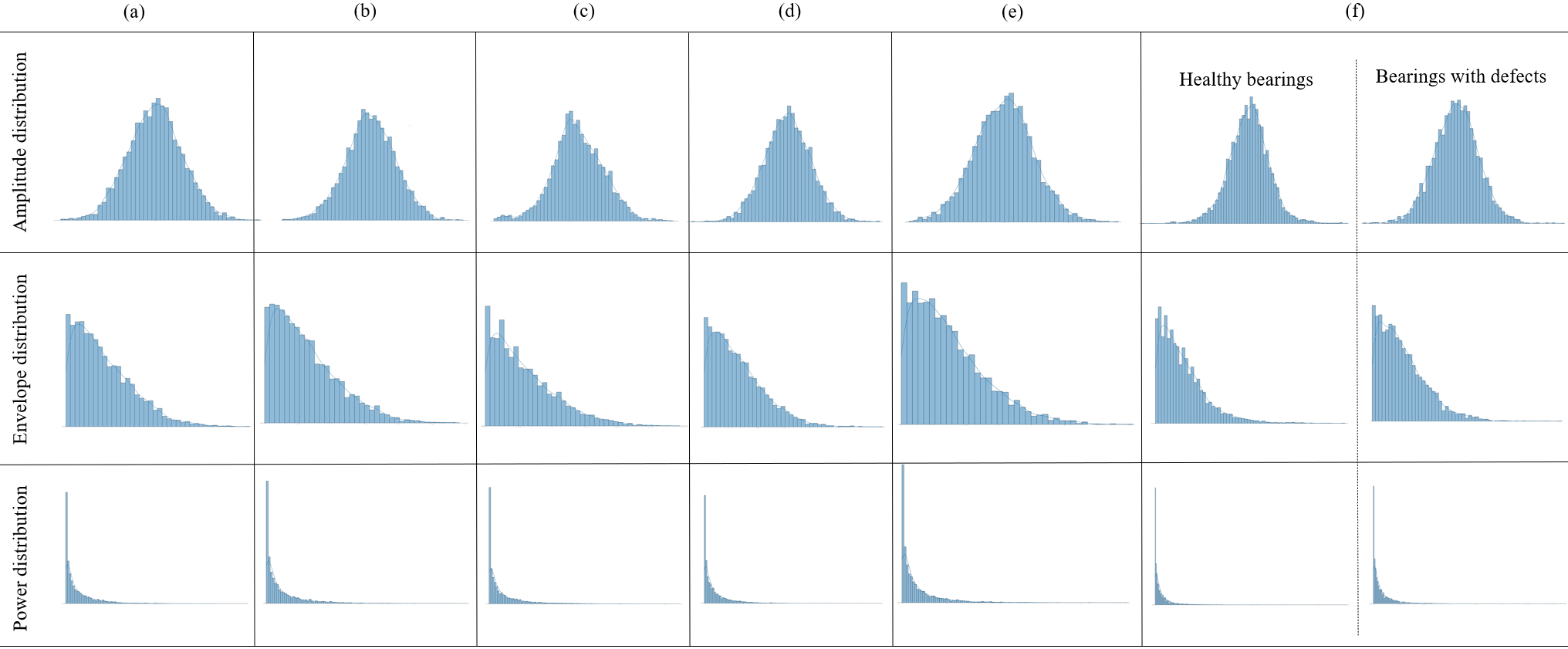}}
\caption{Histograms of sample vibration signals: (a) Gaussian random vibration, (b) acceleration of a flying aircraft, (c) acceleration measurements of a flying UAV, (d) vibration from an SHM setup, (e) vibration generated by a wind turbine gearbox, and (f) vibration generated by rolling bearings.}
\label{pdf}
\end{figure*}

\subsection{Statistical Distribution of Vibration Signals}
The statistical distribution of the vibration samples generated by a VBCM system depends on the characteristics of the monitored process and/or object and is influenced by the surrounding environment. However, when the number of vibration samples $N$ in the signal is large, the signal will approach the Gaussian distribution with a zero mean and a variance of $\sigma^2$  "central-limit theorem." Therefore, a Gaussian random process can accurately model the vibration signal. Accordingly, the signal's envelope $|x(n)|$ follows the one-sided Gaussian distribution, and the power $|x(n)|^2$ has a central chi-square with one degree of freedom. Fig. \ref{pdf} shows histograms of vibration signals generated from (a) Gaussian random vibration, (b) exterior of a flying aircraft \cite{aircraft}, (c) a flying unmanned aerial vehicle (UAV) \cite{uav}, (d) an SHM setup \cite{shm}, (e) a wind turbine gearbox \cite{wind}, and (f) rolling bearings of a rotating machinery\cite{pu}. These vibration sets are chosen to resemble the vibration patterns found in various VBCM applications. They represent healthy or normal vibrations "except vibrations of the rolling bearings (Fig. \ref{pdf}.f), which includes normal and faulty vibrations." It is worth mentioning that an abnormal operation or a failure influences the instantaneous vibration, and hence, it would alter the amplitude distributions of the generated vibration. Although generalizing the aforesaid assumption of Gaussian nature may not be entirely accurate, the histograms presented in Fig. \ref{pdf} show that this assumption would be valid for a broad range of vibration patterns. 

\subsection{Statistical Analysis of the PAPR}

Following the assumption that a vibration signal $x(n)$ follows a Gaussian distribution, its probability density function (PDF) can be expressed as:
\begin{equation}
p(x, \sigma)=\frac{1}{\sigma\sqrt{2\pi}}\times exp\left ( -\frac{x^2}{2\sigma^2} \right)
\end{equation}
Accordingly. the signal's envelope $|x(n)|$  has a one-sided Gaussian distribution; its PDF is given by:
\begin{equation}
p_e(x, \sigma)=\sqrt{\frac{2}{\pi\sigma^2}}\times exp\left (-\frac{x^2}{2\sigma^2} \right ), x \ge 0
\end{equation}
The cumulative distribution function (CDF) of the  signal's envelope is then obtained by:
\begin{equation}
F_e(x,\sigma)=\int_{0}^{x}\frac{\sqrt{2}}{\sigma\sqrt{\pi}}\times exp\left (-\frac{u^2}{2\sigma^2} \right ) du
\end{equation}
using
\begin{equation}
t = \sqrt\frac{u^2}{2\sigma^2}\,
\end{equation}
the CDF can be written as:
\begin{equation}
\begin{split}
F_e(x,\sigma) & = \frac{2}{\sqrt{\pi}} \int_{}^{\sqrt{x^2/2\sigma^2}}exp\left (-t^2 \right ) dt \\
& = erf\left (\sqrt{\frac{x^2}{2\sigma^2}}  \right )
\end{split}
\end{equation}
where $erf\left ( \cdot \right )$ is the error function.
Accordingly, the probability that the signal's power ratio $P =\frac{x^2}{\sigma^2}$ is above a given PAPR threshold $P_o$ can be obtained using the complementary cumulative distribution function (CCDF):
\begin{equation}
\begin{split}
Prob(P>P_o) & =  \text{CCDF} = 1- (\text{CDF})^N \\
& = 1- erf\left (\sqrt{\frac{P_o}{2}}  \right )^N
\end{split}
\label{ccdf}
\end{equation}
where $N$ is the number of samples in the vibration signal $x(n)$.
\begin{figure}[!b]
\includegraphics[width=0.5\textwidth]{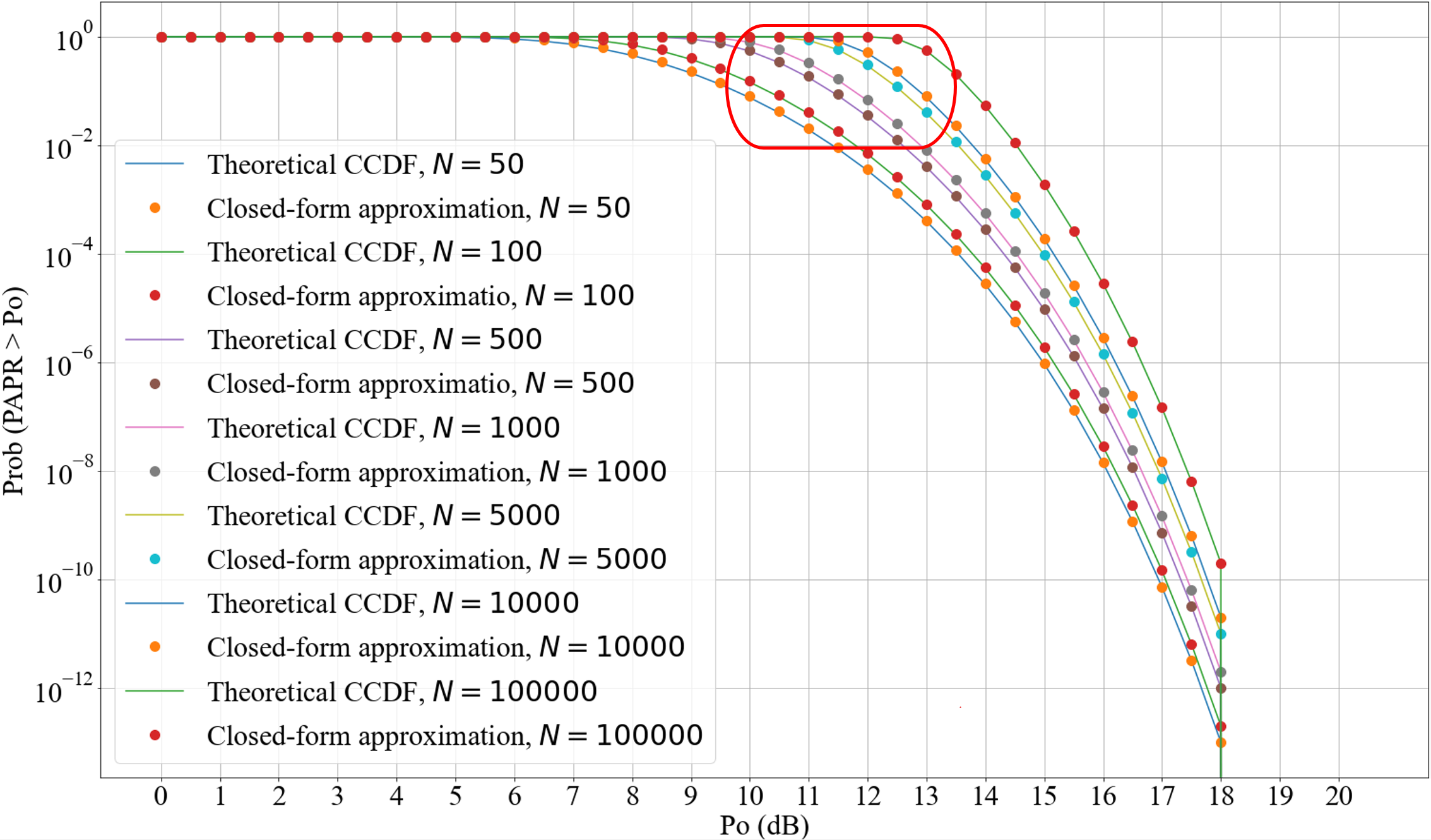}
\caption{Theoretical CCDF and its closed-form approximation for different values of $N$.}
\label{ccdf_th}
\end{figure}
\begin{figure*}[!htb]
\centerline{\includegraphics[width=1\textwidth]{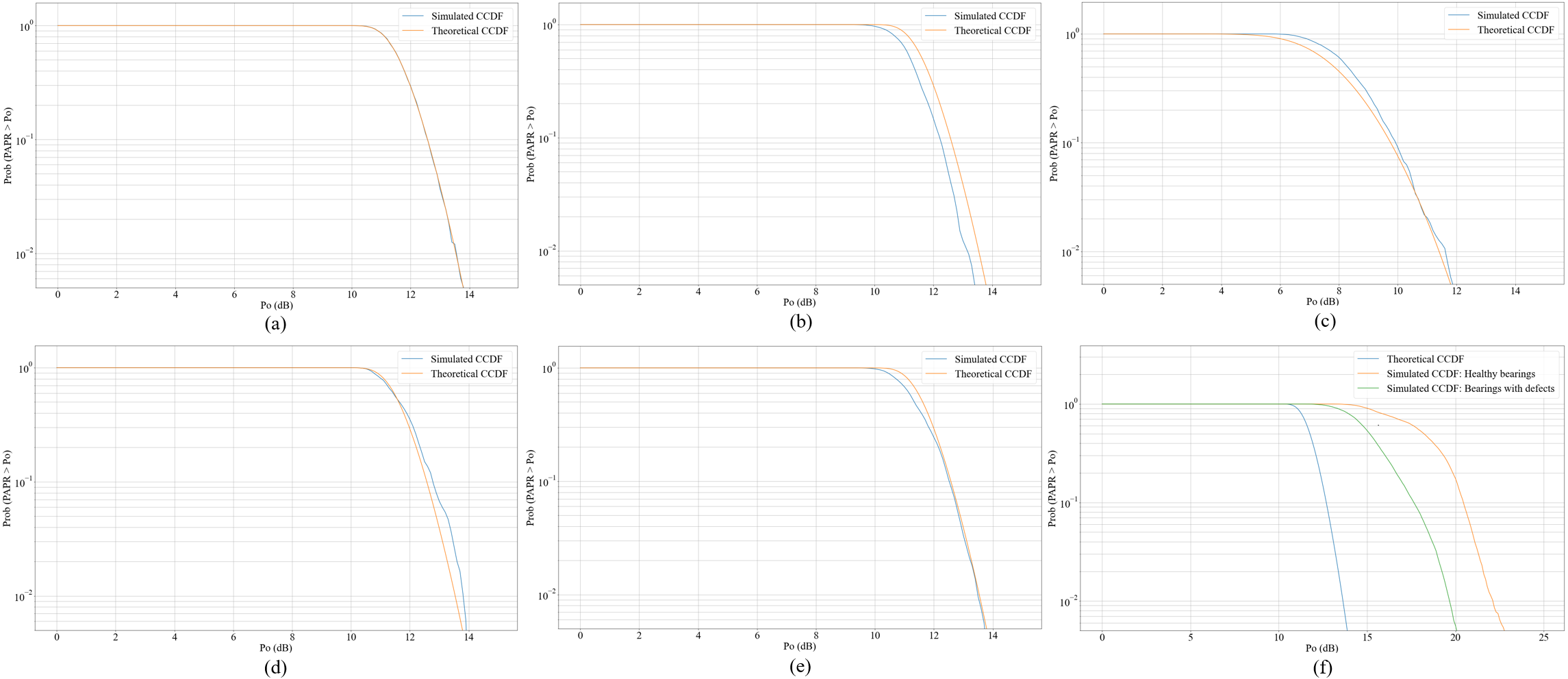}}
\caption{Analytical and simulated CCDFs of (a) Gaussian random vibration ($N=5000$), (b) acceleration of a flying aircraft ($N=5000$), (c) acceleration measurements of a flying UAV ($N=50$), (d) vibration from an SHM setup ($N=5000$), (e) vibration generated by a wind turbine gearbox ($N=5000$), and (f) vibration generated by rolling bearings ($N=5000$).}
\label{ccdf_all}
\end{figure*}
The analytical formula of the CCDF in Eq. \ref{ccdf} is helpful in studying the PAPR of vibration generated in various VBCM systems. Furthermore, the CCDF is a useful metric for evaluating the effectiveness of a PAPR reduction method. Typically, a simulated CCDF is obtained using PAPR-reduced signals and compared to a simulated CCDF of the original signals to evaluate the reduction achieved in the PAPR. A closed-form approximation of Eq. \ref{ccdf} can be obtained using the asymptotic series expansion of the complementary error function $erfc\left(\cdot\right)$:
\newline
Since, 
\begin{equation}
erf\left ( \cdot \right )= 1- erfc\left ( \cdot \right ),
\end{equation}
Eq. \ref{ccdf} can be expressed in terms of $erfc\left ( \cdot \right )$ as follows:
\begin{equation}
Prob(P>P_o)  =  \text{CCDF} = 1- \left( 1- erfc\left (\sqrt{\frac{P_o}{2}}  \right) \right)^N
\label{ccdf2}
\end{equation}
For large values of $\sqrt{P_o/2}$, the complementary error function may be approximated by the asymptotic series expansion:
\begin{equation}
\begin{split}
& erfc\left( \sqrt{\frac{P_o}{2}} \right)  \approx  \frac{e^{-P_o/2}}{\sqrt{\frac{P_o\pi}{2}}} \times\\
& \left( 1-\frac{1}{P_o} + \frac{1\cdot 3}{P_o^2} -  \cdot \cdot \cdot +(-1)^n\frac{(2n-1)!!}{P_o^n}+ \cdot \cdot\cdot \right )\\
\end{split}
\end{equation}
For $ P_o\gg 1$,
\begin{equation}
erfc\left( \sqrt{\frac{P_o}{2}} \right) \approx \frac{e^{-P_o/2}}{\sqrt{\frac{P_o\pi}{2}}}
\label{ccdf3}
\end{equation}
Accordingly, a closed-form approximation of the CCDF can be obtained by substituting Eq. \ref{ccdf3} into Eq. \ref{ccdf2}:
\begin{equation}
Prob(P>P_o)  =  \text{CCDF} = 1- \left( 1- \frac{e^{-P_o/2}}{\sqrt{\frac{P_o\pi}{2}}}  \right)^N
\label{ccdf4}
\end{equation}

Fig. \ref{ccdf_th} shows plots of the CCDF in Eq. \ref{ccdf} and its closed-form approximation in Eq. \ref{ccdf4} for different values of $N$. The plotted CCDF curves show an exact match between the CCDF and its closed-form approximation. Fig. \ref{ccdf_all} shows simulated CCDFs of the aforementioned vibration sets (refer to Fig. \ref{pdf}) along with theoretical CCDFs (Eq. \ref{ccdf4}). It is evident that the simulated CCDFs align with their corresponding theoretical CCDFs, except for the rotating machinery. The mismatch in the case of rotating machinery could be due to the rotating nature of the bearings and speed fluctuations \cite{app5}\cite{mech00}. Additionally, the graphs depicted in Fig. \ref{ccdf_th} and Fig. \ref{ccdf_all} demonstrate that as the number of samples $N$ increases, the likelihood of experiencing a high PAPR increases. Specifically, with the number of samples $N \ge 500$, a PAPR in the range of $10$-$13$ dB is likely to occur. Thus, it can be concluded that vibration signals generally tend to have high PAPR, where peak vibrations that are 10–20 times higher than the average vibrations occur commonly.

\section{Review of signal companding techniques}
To the best of our knowledge, The PAPR of vibration singles and the associated problem of nonlinear power amplification have not been addressed yet in the literature. Nevertheless, this issue has been extensively studied in multi-carrier communications, particularly in orthogonal frequency-division multiplexing (OFDM) systems. In such systems, information is split into parallel streams and carried across orthogonal sub-carriers during each transmission period. These sub-carriers are then summed together to form the transmitted symbol. Transmitting the data over orthogonal sub-carriers helps to reduce interference, minimize the required bandwidth, and increase the data transmission rate. However, the summation of modulated orthogonal sub-carriers would result in very high peaks in the transmitted symbol, which makes the OFDM signal exhibit a high PAPR.  Numerous methods have been suggested in the literature to reduce the high PAPR of OFDM signals to prevent nonlinear distortion and increase the efficiency of the power amplification. These techniques can be broadly categorized into three main categories: Symbol structure modification, peak clipping, and signal companding. Structure modification techniques include block coding \cite{ofdm1}, selective mapping (SLM) \cite{ofdm2}, partial transmission sequence (PTS) \cite{ofdm2}, and tone reservation \cite{ofdm4}. These techniques reduce the PAPR by modifying the structure of the transmitted OFDM symbol. They generally impose restrictions on its parameters and require transmitting side information to reconstruct the symbol at the destination. Therefore, the reduction in PAPR comes at the cost of increased complexity and reduced data rates due to the transmission of side information. Clipping \cite{clip} offers a simple approach to reducing the PAPR by hard-limiting the peaks to a pre-defined threshold. Despite its simplicity, clipping introduces amplitude distortion and spectral spreading. While amplitude distortion is unrecoverable, filtering would reduce spectral spreading. However, the peaks of the filtered-clipped signal could exceed the clipping threshold due to peak power regrowth after filtering. Alternative solutions that help to reduce the clipping distortion involve repeated or iterative clipping \cite{clip} and peak windowing \cite{win}. In contrast to clipping, peak windowing applies soft-limiting to the peaks by multiplying the signal with a window-weighting function. As a result, distortion is reduced since the peaks are smoothly and softly limited. Signal companding is a well-known method in signal processing that involves two steps. First, the signal is transformed into a compressed form at the source. Second, the inverse transform expands the compressed signal at the destination. The compression reduces the signal's dynamic range (DR) and allows for efficient processing of the signal. $\mu$-law and A-law \cite{comp} are the most common companding transforms that are typically applied to speech signals to reduce quantization noise and optimize the required number of bits per sample for analog-to digital conversion. In $\mu$-law and A-law transforms, signal compression is achieved by applying a logarithmic-based transform to enlarge small amplitudes in the signal. Signal companding has gained wide popularity in the OFDM domain compared to other techniques due to its low complexity. Companding has no restrictions on the symbol's parameters and does not require the transmission of side information. Further, companding has better error performance compared to clipping. Using signal companding to reduce the PAPR of OFDM signals was first introduced in \cite{comp1}. However, the scope of analysis was limited to addressing the effect of $\mu$-law companding on the quantization noise. Reducing the PAPR of the signal by applying $\mu$-law companding will increase the signal's average power. This, in turn, improves the signal-to-quantization noise ratio since the small amplitudes are enlarged. However, considering the non-linearity of the HPA, reducing the PAPR by increasing the signal's average power will not prevent the nonlinear distortion since the large peaks are not reduced. In fact, it would lead to more distortion in the signal. The main attention of the ongoing research is directed toward addressing this problem by designing the companding function so that the increase in the signal's average power is avoided \cite{comp2, comp3, comp4, comp5, comp6, comp7}. The published work in this area can be grouped under two main approaches. The first approach involves using additional transforms and/or optimization algorithms, which obviously increases computational complexity. The second approach involves introducing inflexion points in the signal. This allows for independent scaling of large peaks and small amplitudes, which helps to maintain the signal's average power. However, this approach reduces the data rate since the signal's indexes must be transmitted to apply the inverse operations at the destination. A considerable amount of the recent work focuses on utilizing deep learning (DL) models to tackle the problem of PAPR in OFDM \cite{dl, dl2, dl3, dl4, dl5}. DL-based approaches are centered around designing and training the DL models to optimally, sub-optimally, or efficiently learn the function of the corresponding conventional PAPR reduction scheme while mitigating the associated drawbacks.

Choosing the appropriate technique among the options mentioned above for reducing the PAPR of vibration signals starts by understanding the distinctions between OFDM and vibration signals. Regarding structure modification approaches, vibration signals are generated by sensors as raw data, unlike OFDM symbols, which are formed based on a predetermined structure. Therefore, such techniques are not applicable to vibration signals. Clipping introduces unrecoverable distortion in the clipped signal; this can be tolerated in the OFDM signal due to the error correction mechanisms. In VBCM systems, the characteristics of the monitored process/object are described by the waveform and the spectrum of the generated vibration signal. This makes clipping distortion critical and intolerable since it introduces distortion in the generated waveform and alters its spectral contents. Compared to structure modification and clipping, signal companding presents a practical solution for reducing the PAPR of vibration signals without affecting the monitoring process. However, in order to be adopted for VBCM applications, the companding transform should fulfill the following three requirements:
\begin{enumerate}
    \item Avoiding the increase in the signal's average power.
    \item Employ effective signal denoising to eliminate noise from the compressed signal before the expansion stage to mitigate the effects of noise enhancement during the expansion process 
    \item Avoid transmission side information as this will increase the amount of the transmitted data and will lead to more power consumption in the sensor node.
\end{enumerate}
Considering these requirements, the upcoming section introduces the proposed autoencoder-based companding framework.

\section{Signal Companding for Reduction of PAPR}
As stated earlier, signal commanding provides a practical solution for reducing the PAPR of vibration signals without affecting the VBCM process. The key requirements in companding are preventing any increase in the average power of the signal and avoiding the transmission of side information. In this section, we present the lightweight companding-based framework proposed for reducing the PAPR of vibration signals. However, it is convenient first to provide a brief overview of conventional signal companding.

\subsection{Conventional Signal Companding}
The most commonly used type of signal companding is the $\mu$-law companding. its compression function $C(x)$ can be expressed as: 
\begin{equation}
y=C(x)=A \, sgn(x)\,\frac{ln(1+\mu|\frac{x}{A}|)}{ln(1+\mu)}
\end{equation}
where, $x$ is the input signal, $sgn(\cdot)$ is the sign function, $A$ is a normalization constant such that $0<|\frac{x}{A}|<1$, and $\mu$ is the compression parameter. The expansion (inverse) function is expressed as:
\begin{equation}
\begin{split}
& {x}'=  C^{-1}(y)\\
& = A \left [ \frac{\textrm{exp}\left \{  \frac{|y|}{A\, sgn(y)}  ln(1+\mu)\right \}-1}{\mu\,sgn(y)} \right]
\end{split}
\end{equation}
\begin{figure*}[!htb]
\centerline{\includegraphics[width=1\textwidth]{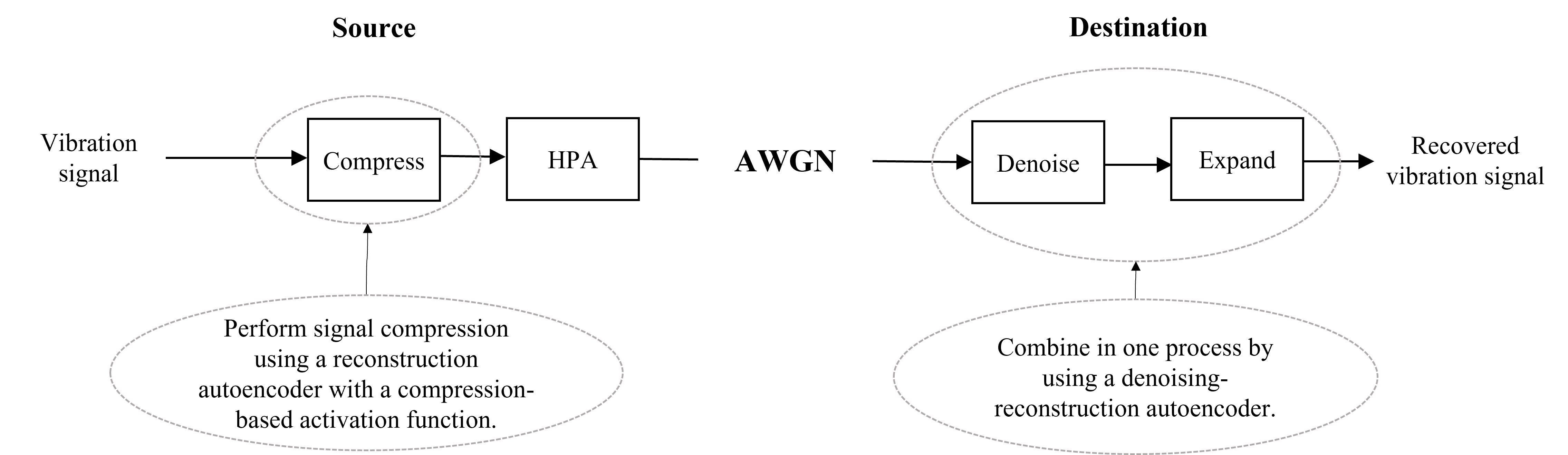}}
\caption{Proposed framework for autoencoder-based companding of vibration signals.}
\label{fw}
\end{figure*}
 Fig. \ref{mu}.a displays the compression profile of $\mu$-law for different values of compression parameter $\mu$. Increasing the $\mu$ value leads to more enlargements of small amplitudes, resulting in a higher average power of the signal. Hence, the signal's average power increases as a function of $\mu$ as illustrated in Fig. \ref{mu}.b. Since the signal's peaks are maintained unchanged in $\mu$-law companding, the reduction in the PAPR of the signal is achieved solely by increasing its average power. However, to avoid nonlinear distortion in the vibration signal and improve the power efficiency of the HPA, it is required to reduce the PAPR by reducing the signal's peaks instead of increasing its small amplitudes. In other words, it is required to reduce the PAPR of the signal while avoiding any increase in its average power. Another issue with conventional companding is the undesired effect of enhancing the accumulated noise at the destination due to expansion operation \cite{comp1}-\cite{comp6}. Thus, it is crucial to reduce the effects of noise enhancement by applying effective denoising to the compressed signal prior to expanding it.  
\begin{figure*}[!htb]
\centerline{\includegraphics[width=1\textwidth]{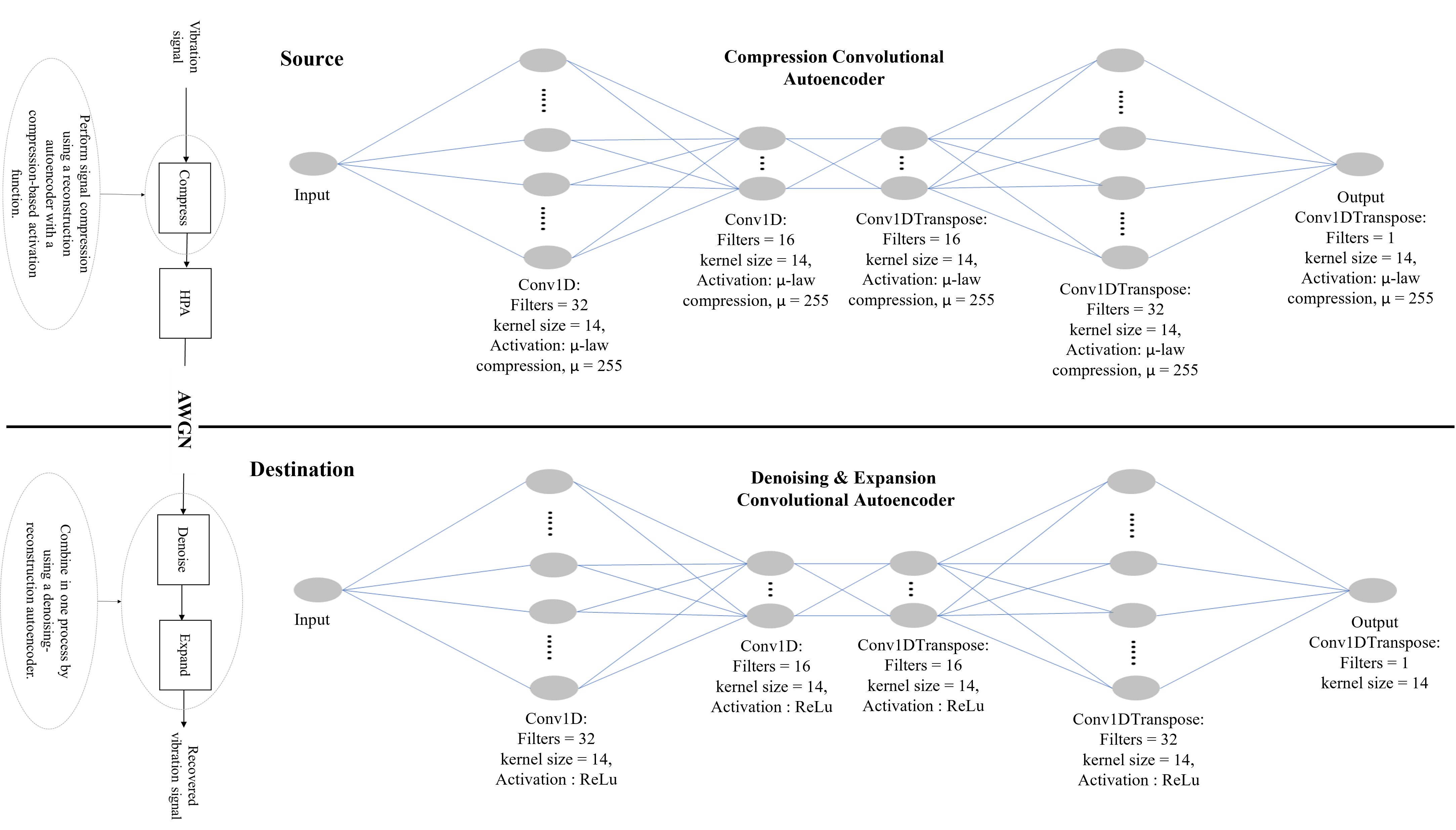}}
\caption{Structures of the reconstruction autoencoders used in the proposed framework.}
\label{setup}
\end{figure*}
\subsection{Proposed Signal Companding}
Here, we introduce our proposed framework for efficient companding of vibration signals that addresses the aforementioned issues of conventional companding. The main aspects of the framework are illustrated in Fig. \ref{fw}. Specifically, signal compression at the source is achieved using a lightweight reconstruction autoencoder with a compression-based activation function. At the destination, signal denoising and expansion operations are combined in one process by using a denoising-reconstruction autoencoder.

\subsubsection{Signal Compression at the Source} 
In the first place, the raw vibration signals are first smoothed to remove measurement noise. Then, a reconstruction autoencoder is trained to learn the smoothing function and reconstruct these smoothed signals (target signals). By using a compression-based activation function in the autoencoder layers, the autoencoder will learn how to reconstruct the input signals based on the target signal and, at the same time, compresses the learned presentations of the input signal. As a result, the output of the trained autoencoder is a smoothed and compressed version of the input signal.
\newline
\textbf{Average power of the compressed signal:} During the training process, the autoencoder compresses the signal’s representations in each layer while, at the same time, it learns to minimize the loss between the input signal and the target (smoothed) signal. Here, the average power of the target signal represents an upper bound on the average power of the reconstructed signal. This will avoid any increase in the average power of the output (smoothed-compressed) signal. Further, the joint mechanism of reconstruction-compression will maintain the average power of the output signal as close as possible to the average power of the target signal.
\newline
\textbf{Compression loss as a lower bound on the training loss:} To efficiently train the source autoencoder, it is important to consider the following factors:
\begin{itemize}
\item The training objective of the source autoencoder is to reconstruct a smoothed and compressed version of the input signal rather than reconstructing the signal in its original form.
\item The compression-based activation function in the autoencoder layers implies that there will be a compression loss or a minimum error floor between the output signal and the target signal caused by the compression mechanism. 
\end{itemize}
Considering these factors, it is not required to minimize the loss until maximum convergence. Instead, it is desirable to train the autoencoder until reaching this error floor, which is determined by compression loss. Theoretically, the compression loss is the difference between the reconstructed signal and its "perfectly reconstructed" compressed form. Mathematically, the compression loss ($CL$) can be calculated as the difference between the target signal $x$ and its compressed and power-preserved form $x_{pc}$:
\begin{equation}
\begin{split}
CL = error(x,x_{pc}), \\
 x_{pc} = P\left ( AF(x) \right ),
\end{split}
\label{cl}
\end{equation}
where $AF$ is the compress-based activation function, and $P$ is the power scaling operation. The $error$ function can be either the mean absolute error (MAE) or the mean squared error (MSE). To efficiently train the autoencoder, the compression loss can be utilized to set a baseline for the loss during the training. It can be empirically determined using the input and the target training signals to obtain the average $CL$ according to Eq. \ref{cl}. Generally, the effect of compressing a signal and preserving its average power can be approximated by applying a hard limiter to the signal with a peak-limiting threshold equal to the maximum peak of its compressed form. Accordingly, the clipping noise, which is the power of the clipped portion, can be used as an estimate of the compression loss. Given a target PARR ($PAPR_{t}$), the maximum peak $Peak_{c}$ of the compressed and power preserved signal $x$ equals to:
\begin{equation}
Peak_{c} = \sqrt{PAPR_{t} \times P_{in}},
\end{equation}
where $P_{in}$ is average power of the signal. Following the assumption of the Gaussian nature of $x$, it can be modeled as a Gaussian random process with a zero mean and a variance $\sigma^2 = P_{in}$. Thus, the probability that, at any given time, the signal $x$ takes the value  $Peak_{c}$ is given by:
\begin{equation}
\begin{split}
& Prob\left \{ x(t) = Peak_{c}\right \}\\
& =p(x)=\frac{1}{\sqrt{2\pi P_{in}}}\times\text{exp}\left ( -\frac{x^2}{2P_{in}} \right )
\end{split}
\end{equation}
Since  maximum peak of the signal is limited to $Peak_{c}$, the clipping noise ($CN$) is given
by:
\begin{equation}
CN = 2\int_{Peak_{c}}^{\infty}\left ( x-Peak_{c} \right )^2p(x)dx 
\end{equation}
Using the analysis presented in \cite{pa}, $CN$ can be approximated as:
\begin{equation}
CN\cong 2\sqrt{\frac{2}{\pi}}\times\sigma^2 \times(\sqrt{PAPR_{t}})^{-3}\times\text{exp}\left ( -\frac{PAPR_{t}}{2} \right)
\end{equation}
\begin{table}[!b]
  \caption{Comparison between the proposed method and conventional training}
  \label{compare}
  \includegraphics[width=0.5\textwidth]{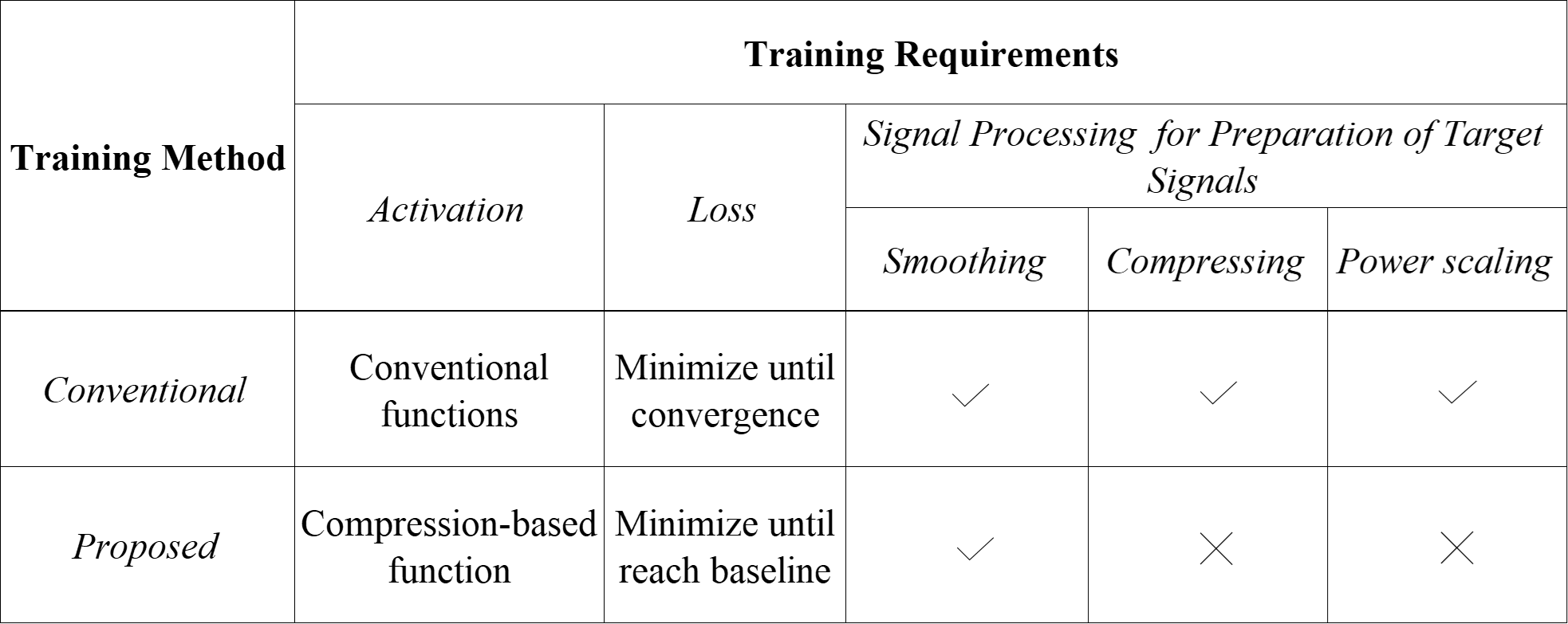}
\end{table}
An alternative way for training the source autoencoder involves utilizing conventional activation functions like the rectified linear unit (ReLU) and using compressed and power-reserved forms of the input signals as the training targets. However, the proposed method, in contrast, requires less training time and involves less signal processing, as highlighted in Table \ref{compare}, which compares the proposed method and conventional autoencoder training. The comparison demonstrates the lightweight nature of the proposed framework, which requires fewer computations and reduced signal processing, making it more power-efficient and less complex than conventional methods.

\subsubsection{Signal Denoising and Expansion at the Destination} 
In order to train the denoising-expansion autoencoder at the destination, the vibration signals of the training set are firstly compressed using the compression function $AF$. Then, they are corrupted with AWGN noise at a desired signal-to-noise ratio (SNR). The autoencoder is then trained with the noisy, compressed signals as the input and the original signals as the target. By training the autoencoder to minimize the loss between input and target signals, it will learn the expansion mechanism. Additionally, the autoencoder weights will be tuned to remove the noise.  Hence, it simultaneously acts as an expanding function and a denoising filter.
\begin{figure}[!htb]
\centerline{\includegraphics[width=0.5\textwidth]{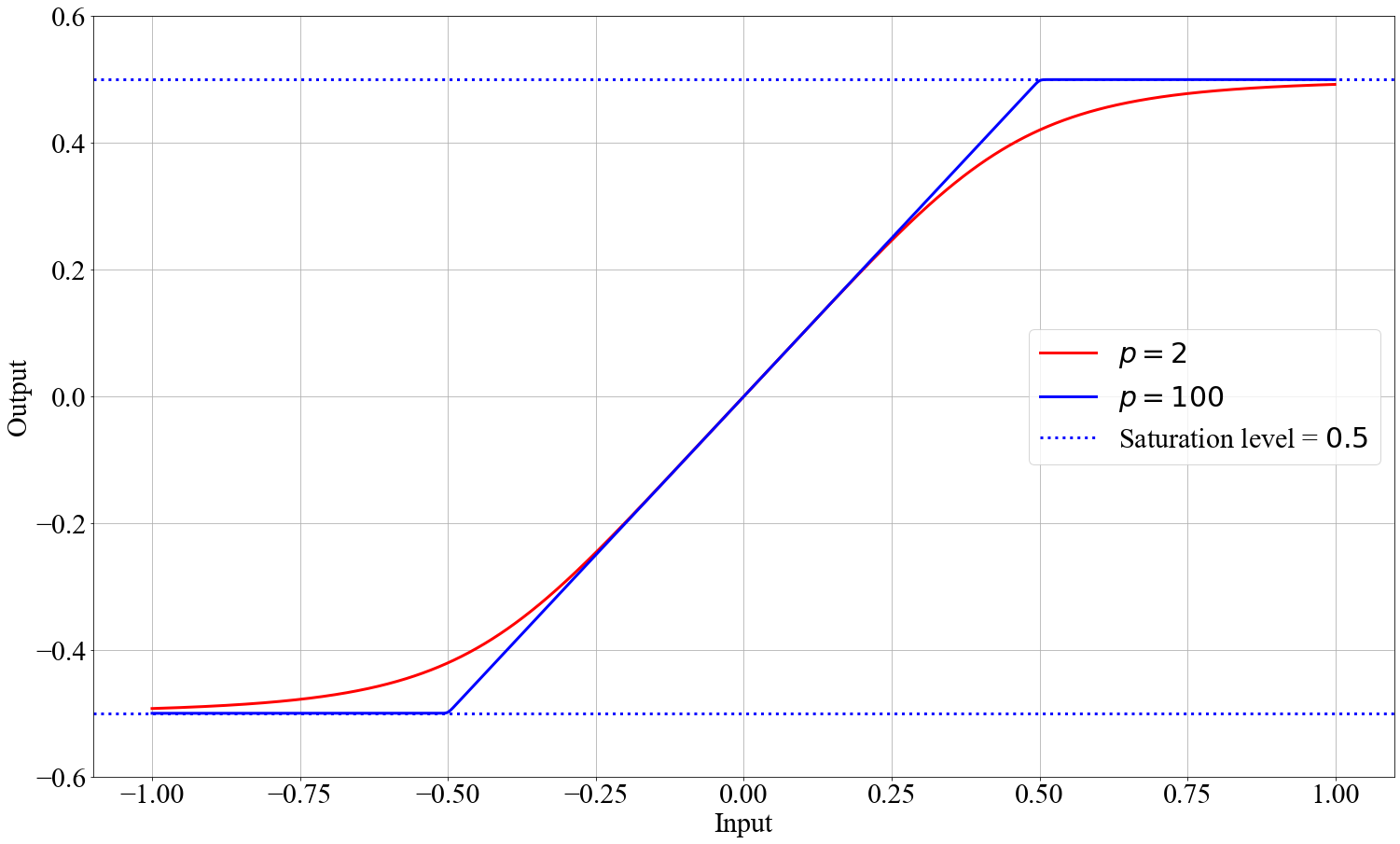}}
\caption{AM-AM and AM-PM conversions of the Rapp SSPA model.}
\label{hpa}
\end{figure}

\section{Modeling Nonlinear Power Amplification}

A practical HPA has a limited linear range and exhibits a nonlinear behavior at its saturation point, as deposited in Fig. \ref{fig7.1.2}. As previously stated, to achieve linear amplification of the signal and avoid nonlinear distortion, it is essential that the peaks of the signal remain within the linear range of the HPA. Otherwise, a large IBO should be used, which as mentioned earlier, reduces the HPA efficiency. Reliable modeling of the HPA is crucial for the accurate evaluation of nonlinear power amplification effects on the signal. A power amplifier is typically modeled by its amplitude-to-amplitude (AM/AM) and amplitude-to-phase (AM/PM) conversion functions. The (AM/AM) conversion is used to characterize the amplitude distortion, which is the relationship between the input power (amplitude) and the output power (amplitude). (AM/PM) conversion is used to characterize phase deviation (distortion) caused by amplitude variations. A widely accepted solid-state power amplifier (SSPA) model is the Rapp model \cite{rapp}. It has a frequency-nonselective response with a smooth transition from linearity to saturation as input amplitude approaches the saturation level. Its (AM/AM) conversion function is:
\begin{equation}
\begin{split}
A_{out} & = a\frac{A_{in}}{\left ( 1+\left [ \left ( \frac{aA_{in}}{A_{sat}}\right )^2 \right ]^p \right)^{1/2p}}\\
& \text{with,} \, A_{sat}\geq 0, \, a\geq0, \, \text{and}\, p\geq0
\end{split}
\end{equation}
where $A_{in}$ is the input amplitude, $A_{sat}$ is the saturation level, $a$ is the gain, and $p$ is a positive number to control the nonlinearity characteristics of the HPA. The (AM/PM) conversion of the SSPA  is small enough and can be neglected \cite{rapp}. Fig. \ref{hpa} shows the (AM/AM) conversion curve of the model with different values of $p$. As it is shown, as the value of $p$ increases, the model converges to a hard limiting amplifier. For large values, the model becomes precisely linear until it reaches its output saturation level. A good approximation of existing amplifiers is obtained by choosing $p$ to be in the range of $2$ to $3$ \cite{rvano}. In this paper, the Rapp model with $p=2$ and $a=1$ is used to simulate the nonlinear power amplification of vibration signals.
\begin{figure*}[!htb]
\centerline{\includegraphics[width=1\textwidth]{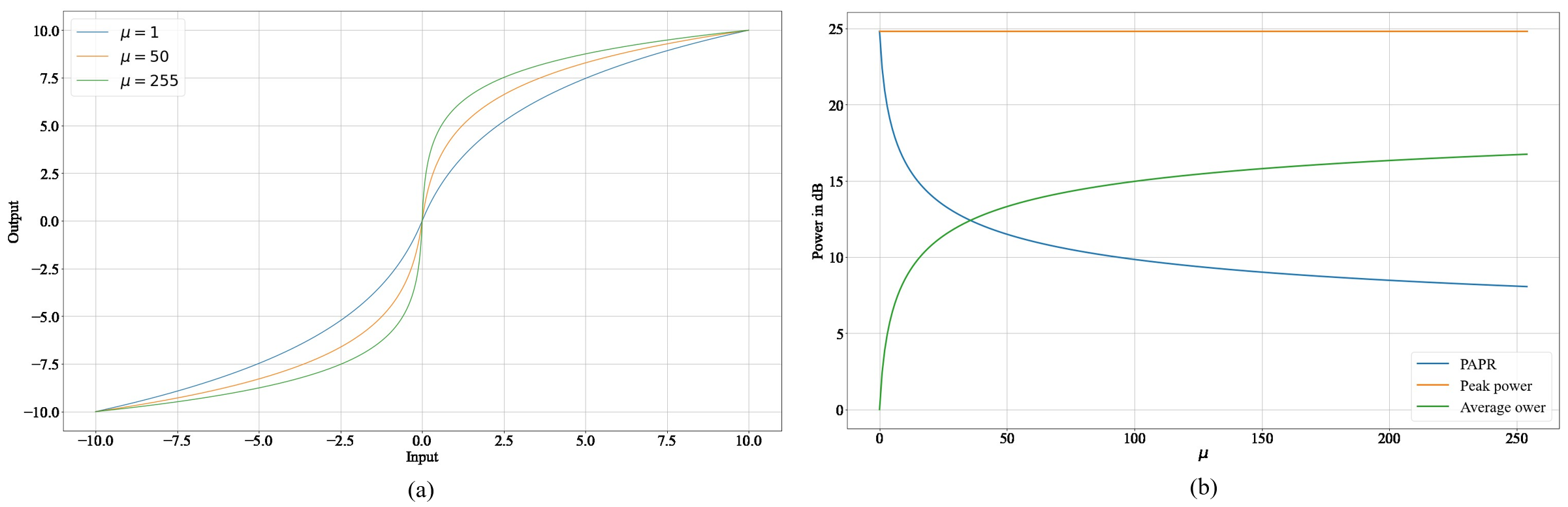}}
\caption{$\mu$-law: (a) compression profile with different values of $\mu$, and (b) peak and average powers as a function of $\mu$ .}
\label{mu}
\end{figure*}

\section{Performance Evaluation}

The vibration signals of the Paderborn University (PU) bearing dataset \cite{pu} (Vibration set (f) in Section III) are used to demonstrate the effectiveness of the proposed framework in reducing the PAPR and mitigating the effects of HPA nonlinearity. This dataset is selected because it includes actual vibration signals from a real system during both healthy and faulty operations. The fault types include Inner Race (IR) defects, Outer Race (OR) defects, and combined defects. More information about the PU dataset can be found in \cite{mech1}. To create the training and testing sets, the vibration measurements of the PU dataset are segmented into segments of $0.1$ seconds ($6,400$ data points). This results in $16,005$ vibration signals in total. Accordingly, the dataset is split into $11,202$ samples for training ($70\%$) and $4,803$ samples for testing ($30\%$). Adam optimizer (learning rate = $0.001$) is used to train the autoencoders and minimize the MSE. The framework is implemented using Python, Keras library \cite{keras}, TensorFlow \cite{tf}, and SciPy library \cite{scipy}.

\subsection{Experimental Setup}
One-dimensional convolutional (Conv1D) layers are used to implement the autoencoders. The structures and the parameters of the autoencoders are shown in Fig. \ref{setup}. The activation function ($AF$) used in the source autoencoder is based on the $\mu$-law compression and expressed as:
\begin{equation}
\textit{AF} = sgn(x)\,\frac{ln(1+255\times|x|)}{ln(1+255)}
\label{act}
\end{equation}
\subsubsection{Training the Source Autoencoder}
To train the source autoencoder, the raw signals of the training set are smoothed in the first place. Then, the autoencoder is trained using Eq. \ref{act} as the activation function, with the raw signals as the inputs and the smoothed signals as the targets.

\subsubsection{Training the destination autoencoder}
Two training scenarios are considered for the destination autoencoder:
\paragraph{Noise-free scenario} The destination autoencoder is trained in reverse order compared to the source autoencoder. First, the signals of the training set (training targets of the destination autoencoder) are used as inputs to the source autoencoder \textbf{\textit{that has already been trained}}. Subsequently, the compressed-smoothed signals obtained from the source autoencoder are employed as the training input for the destination autoencoder.
\paragraph{Noisy scenario} To count for the accumulated noise in practical situations, the obtained (smoothed-compressed) signals from the source autoencoder are randomly and equally corrupted with a zero-mean AWGN of $-5$ dB and $0$ dB SNR levels, and the destination autoencoder is trained accordingly.
\begin{figure}[!htb]
\includegraphics[width=0.5\textwidth]{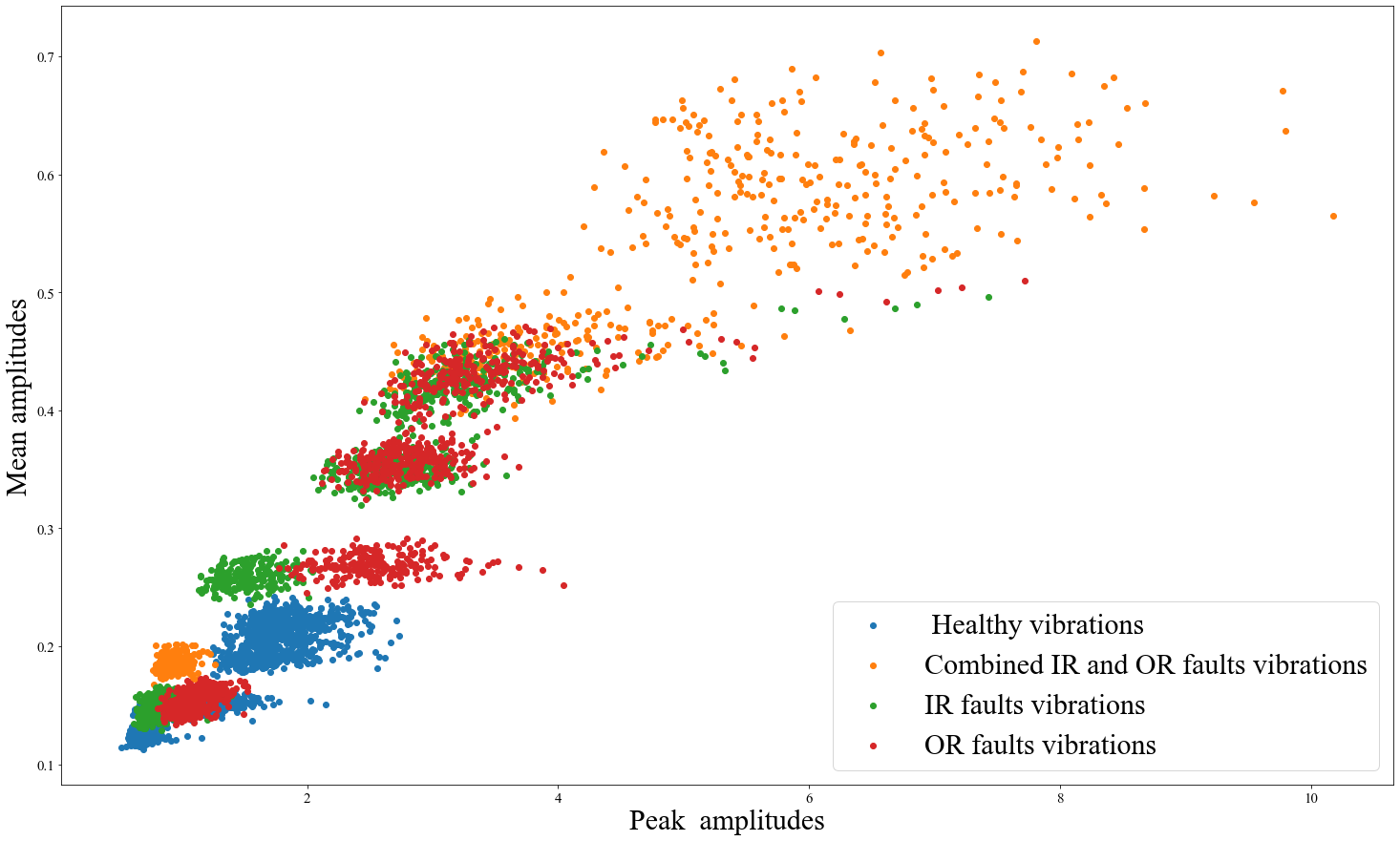}
\caption{Signal's amplitude constellation of the PU dataset.}
\label{sp_pu}
\end{figure}

\subsection{Performance Metrics}
To show the effectiveness of the proposed framework, its performance is evaluated in the presence of nonlinear power amplification and AWGN against the cases of $\mu$-law compression and no compression. The saturation level ($A_{sat}$) of the HPA is set to the mean average power of the original (uncompressed) vibration signals. The performance of the proposed framework is evaluated in terms of the following five aspects:
\subsubsection{\textbf{PAPR reduction}} The CCDF is used to measure the PAPR reduction capability of the proposed framework.
\subsubsection{\textbf{Average power of compressed signals}} This will assess the increase in the average power resulting from the compression.
\subsubsection{\textbf{Distortion due to the nonlinear power amplification}} To assess the nonlinear distortion and HPA power efficiency, we adapt the concepts of \textit{\textbf{amplitude constellation of vibration signals and Error Vector Magnitude (EVM)}}. Signal constellation diagrams and EVM are widely used in telecommunications systems to represent modulated signals and evaluate system-level performance.\
\newline
\textbf{Amplitude constellation of vibration signals:} To obtain the amplitude constellation of a given set of vibration signals, each signal $s_i$ is expressed in terms of its peak and mean amplitudes as follows:
\begin{equation}
s_i = (A_{i_{p}}, A_{i_{m}}), i = 0,...., v-1,
\end{equation}
where $A_{i_{p}}$ is the signal's peak amplitude, $A_{i_{m}}$ is the signal's, mean amplitude, and $v$ is the number of vibration signals in the set. Accordingly, the constellation can be displayed as a scatter plot on the $x-y$ plane where $x$ and $y$ represent the signal peak amplitudes $A_{i_{p}}$ and average amplitudes $A_{i_{m}}$, respectively. Fig. \ref{sp_pu} displays the amplitude constellation of the PU dataset; the points are color-coded according to their health condition so that the signals of the same health condition share the same color. The position of a given signal in the constellation indicates both its peak and average amplitude and the distance--- in terms of these amplitudes--- between the signal and the other signals in the constellation. As mentioned previously,  amplitudes of vibration signals are directly related to the monitored process and/or object. Therefore, the constellation can offer crucial information and insight into the health of the monitored system. Further, as the distortion impacts both the peak and average amplitudes of the signal, its position in the constellation will be altered accordingly. This offers the opportunity to visually evaluate the nonlinear distortion by comparing the amplitude constellation of the amplified-expanded test vibration signals, denoted as $con_{amp}$ to the reference constellation of the original test signals, denoted as $con_{ref}$.
\newline
\textbf{Error Vector Magnitude (EVM):} The EVM can be utilized to quantify nonlinear distortion, assess HPA efficiency, and evaluate the effectiveness of the signal companding scheme. To obtain the EVM, the error vectors of the amplified-expanded test signals with respect to their reference test signals are first calculated from the corresponding constellations $con_{amp}$ and $con_{ref}$. The error vector $error_v$  between two points $s_1 = (A_{1_{p}}, A_{1_{m}})$ and $s_2 = (A_{2_{p}}, A_{2_{m}})$ on the constellation is given by:
\begin{equation}
\begin{split}
error_v = [err_p, err_m], \\
 err_p = A_{1_{p}} - A_{2_{p}},\\
 err_m = A_{1_{m}} - A_{2_{m}}
\end{split}
\end{equation}
Accordingly, the EVM can be calculated as the mean or the RMS value of the magnitudes of these obtained error vectors. It can be expressed as:
\begin{equation}
EVM = \frac{\sqrt{\frac{1}{V}\sum_{i=0}^{V-1}\left | error_v[i] \right |^2}}{\textit{EVM Normalization Reference}} \times 100
\end{equation}
 where $V$ is the number of vibration signals in the test set and $\left|error_v[i] \right|$ is the magnitude of the $i$-th  error vector. In the above equation, EVM is normalized by $\textit{EVM Normalization Reference}$, which equals the maximum magnitude in the reference constellation $con_{ref}$. Hence, the EVM quantifies the power loss and amplitude distortion caused by HPA nonlinearity. In practical situations, the EVM quantifies the combined impact of all signal impairments within a VBCM system (such as distortion and noise effects), enabling measuring the overall system degradation using a single value.

\subsubsection{\textbf{Spectral spreading}} Spectral spreading or spectral broadening refers to situations when a signal's spectrum becomes wider due to nonlinear processing, such as logarithmic-based compression and nonlinear power amplification. Nonlinearity imposed on the signal's envelope causes an undesirable increase in the power of the side lopes of the power spectral density (PSD). This makes PSD an appropriate measure of spectral regrowth. Accordingly, we use the mean PSD to evaluate the compressed-amplified signals' spectral spreading. Welch's overlapped segment averaging method \cite{welch} is used to estimate the PSD.  The method involves segmenting the signal using a moving window and computing each segment's fast Fourier transform (FFT). The PSD is then estimated as the average of the computed FFTs over all segments. In this paper, we use the following settings for Welch's PSD estimation:
\begin{itemize}
\item  Window: Hamming window of a length equals to $N/2$, where $N$ is the length of the vibration signal. This length is selected to obtain a PSD with a good resolution since reducing the window length would affect the resolution.
\item Overlap between segments: $50\%$ overlap. With a window length of $N/2$, an overlap of $50\%$ results in a total of 3 segments, reducing the averaging-error variance compared to using two segments only and, simultaneously, avoiding introducing a high correlation between the segments.   
\item Number of discrete Fourier transform points (NFFT): $NFFT = 8192$. This is calculated using the conventional method where NFFT is set to be equal to $2^p$, where $p$ is the smallest power of $2$ that is greater than or equal to $N$; which in this case equals $13$.
\end{itemize}

\begin{figure*}[!htb]
\centerline{\includegraphics[width= 1\textwidth]{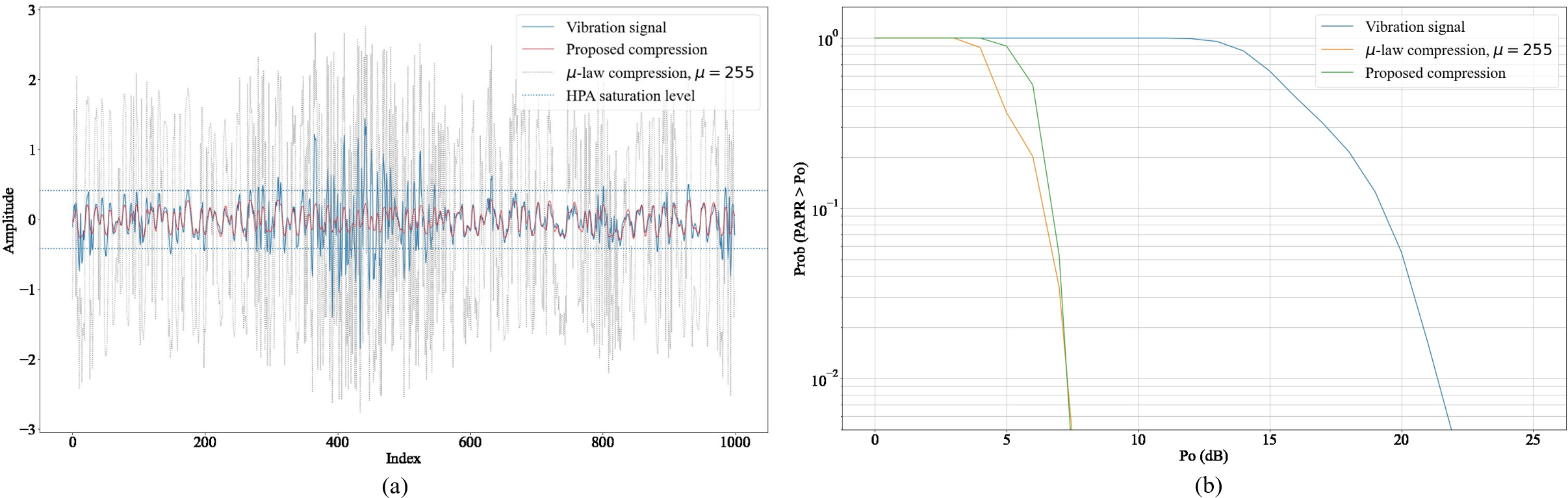}}
\caption{(a) Original and compressed vibration signals and (b) CCDFs of original and compressed signals.}
\label{comp_exp_pm}
\end{figure*}

\subsubsection{\textbf{Signal denoising}} The SNR of the vibration signals after denoising, denoted as $SNR_{d}$ is used to assess the effectiveness of the proposed framework in reducing noise. $SNR_{d}$ is expressed as:
\begin{equation}
SNR_{d}\text{ (dB)}= 10\times \text{log}_{10}\left ( \frac{\sum_{n=0}^{N-1}\left | x(n) \right |^2}{\sum_{n=0}^{N-1}\left (| x(n)-x'(n) \right |^2)} \right )
\end{equation}
 where, 
\newline $x$:  is the original vibration signal,
\newline $x'$: is the denoised signal,
\newline $N$: is the length of the vibration signal.

\begin{figure}[!htb]
\includegraphics[width=0.5\textwidth]{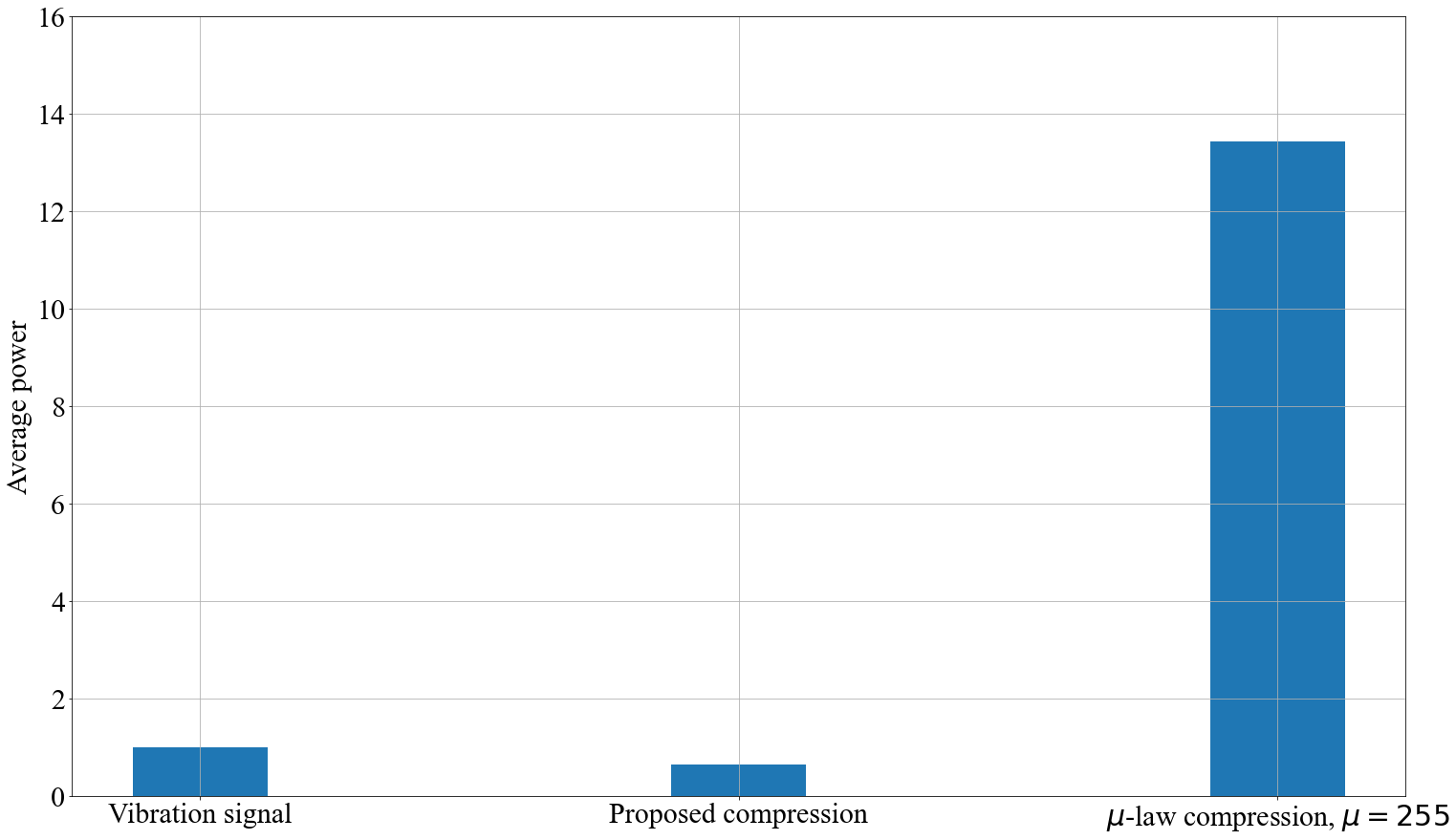}
\caption{Average power ratios of original and compressed signals with respect to normalized average power of original signals.}
\label{pwr}
\end{figure}
\begin{figure*}[!htb]
\centerline{\includegraphics[width=0.8\textwidth]{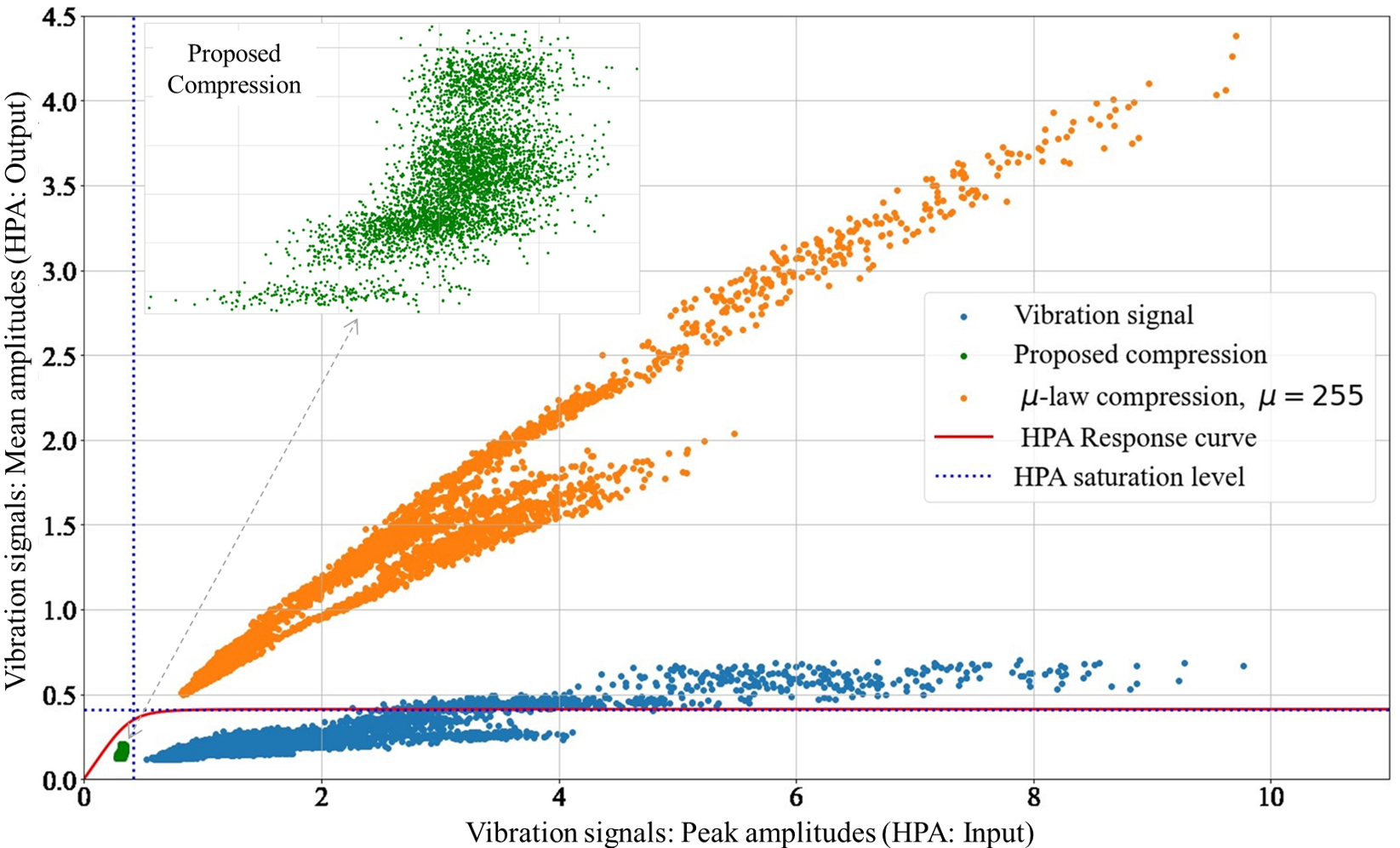}}
\caption{Constellations of uncompressed and compressed signals along with HPA response curve.}
\label{sp_hpa}
\end{figure*}
\begin{figure}[!htb]
\includegraphics[width=0.5\textwidth]{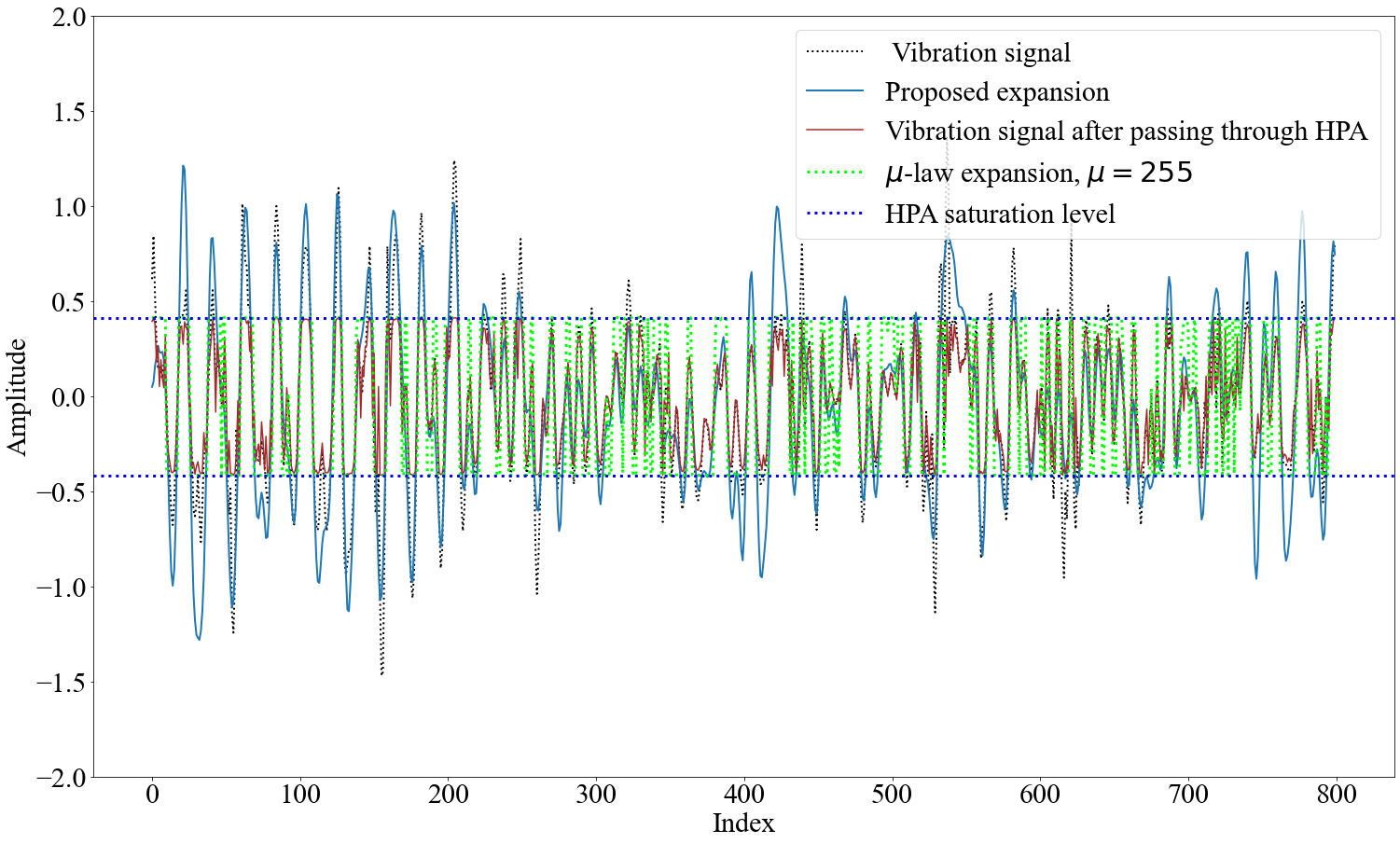}
\caption{Original and expanded vibration signals after passing through HPA.}
\label{exp}
\end{figure}
\begin{figure*}[!htb]
\centerline{\includegraphics[width=1\textwidth]{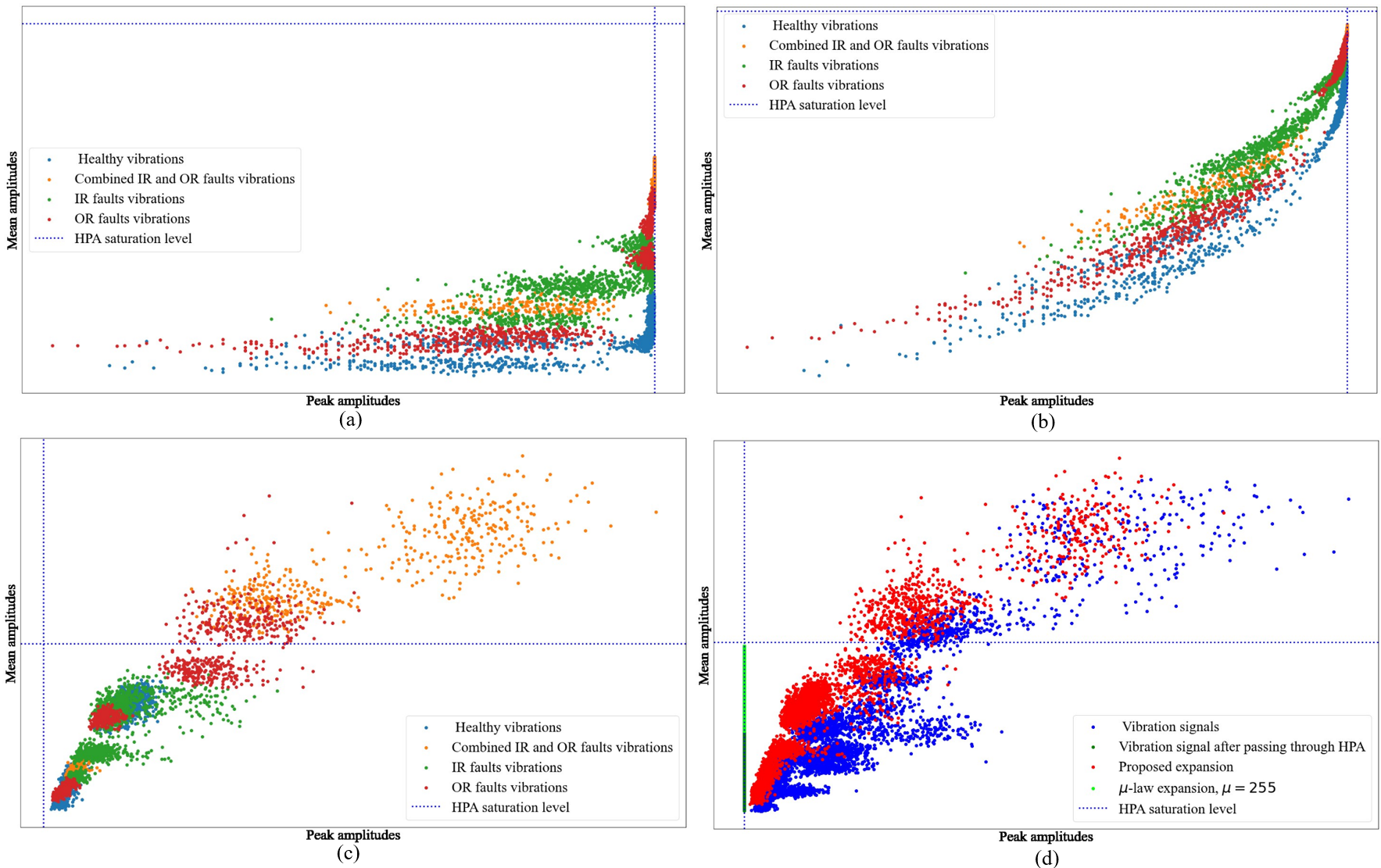}}
\caption{Amplitude constellations after passing through the HPA: (a) uncompressed signals, b) $\mu$-law expanded, (c) proposed-expanded signals, and (d) original signals of the PU dataset along with uncompressed and expanded signals.}
\label{sp_pm_all}
\end{figure*}
\begin{figure}[!htb]
\includegraphics[width=0.5\textwidth]{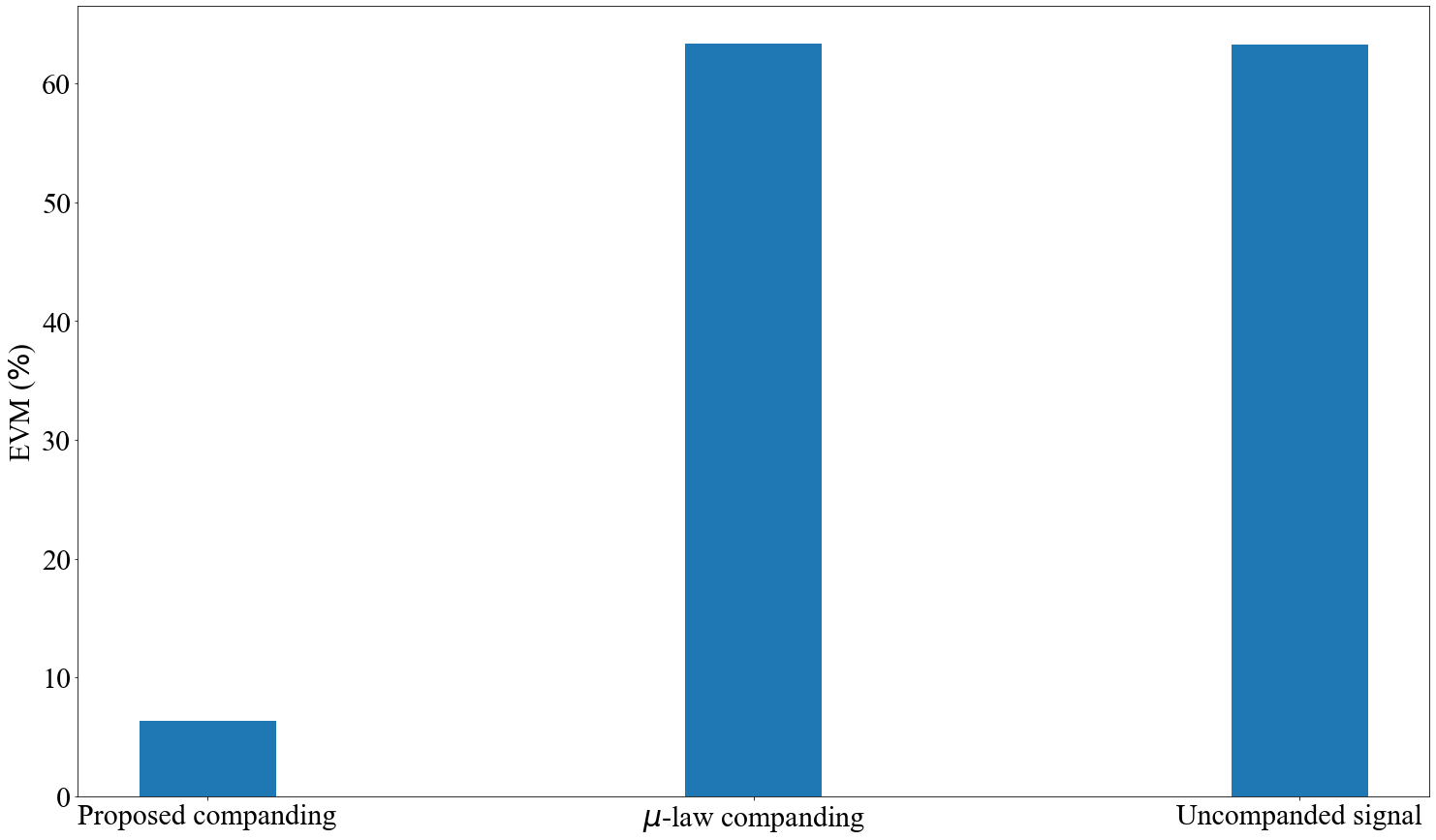}
\caption{EVM values of uncompressed and expanded signals.}
\label{evm}
\end{figure}
\begin{figure}[!htb]
\includegraphics[width=0.5\textwidth]{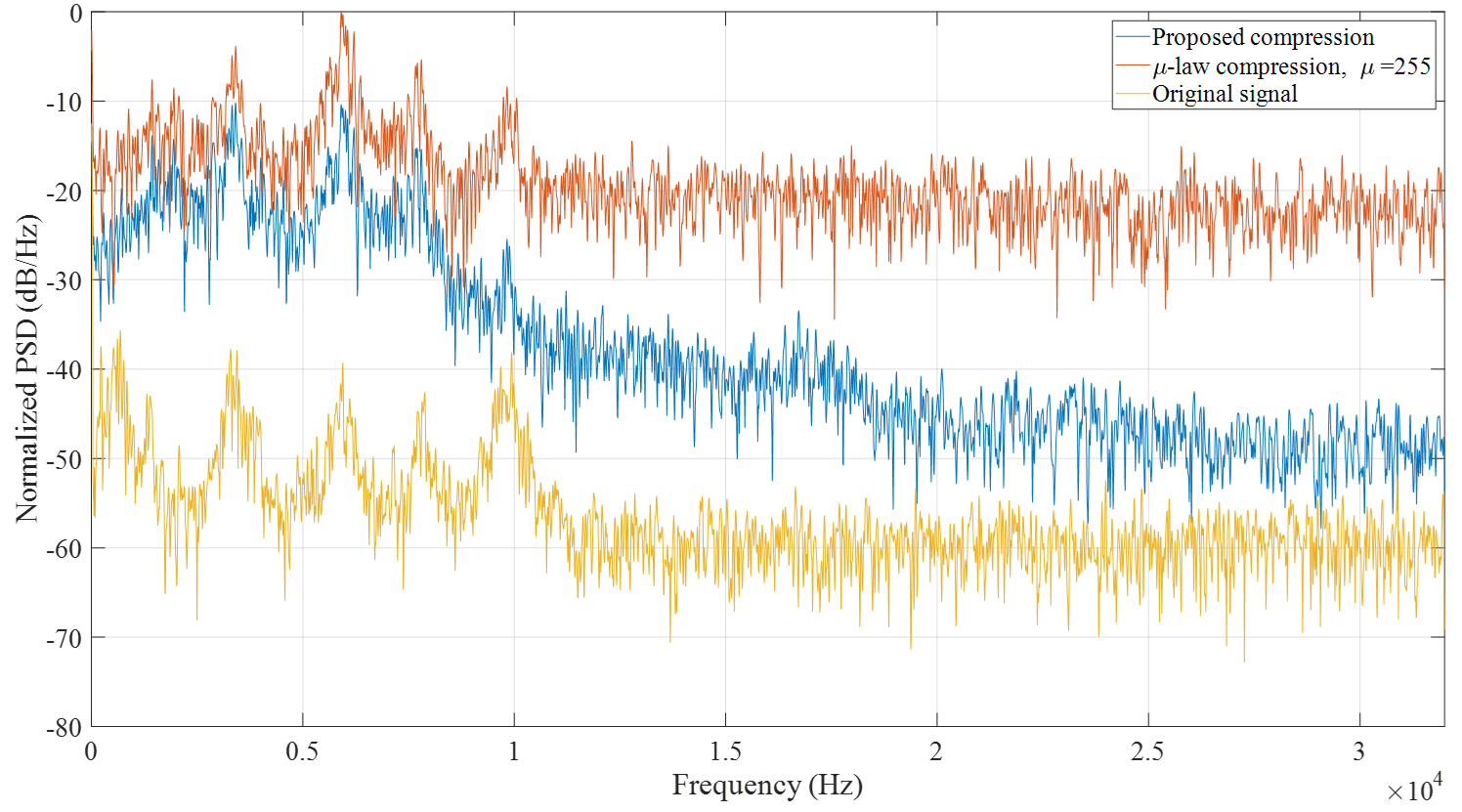}
\caption{Normalized mean PSD plots of original signals, proposed-compressed signals, and $\mu$-law compressed signals after passing through the HPA.}
\label{psd}
\end{figure}

\section{Results and Discussion}
Before presenting and discussing the obtained results, it is convenient to demonstrate the important role of proper signal companding in mitigating the effects of nonlinear power amplification. In Fig. \ref{comp_exp_pm}.a, it can be seen that the uncompressed vibration signal experiences high-amplitude excitations between the signal's indices $350$ and $550$. These amplitude excitations exceed the HPA saturation level, plotted as a dashed horizontal line in the Figure. In the absence of a proper signal companding mechanism, such excitations in the signal's waveform- directly related to the monitored system and would carry vital information about its current condition- are subject to the nonlinear distortion of the HPA.
\subsubsection{\textbf{PAPR reduction}} As shown in Fig. \ref{comp_exp_pm}.b, both $\mu$-law compression, and the proposed compression are effective in reducing the PAPR of the test vibration signals. Specifically, the proposed compression and $\mu$-law compression have reduced the probability of exhibiting a PAPR of $8$ dB in the test vibration signals from $100\%$ to $0.02\%$ and $0.06\%$, respectively. However, as previously mentioned, $\mu$-law compression relies on preserving the signal's peak amplitude while increasing its small amplitudes. Consequently, all amplitudes in the $\mu$-law compressed form of the vibration signal surpass the HPA saturation level, as illustrated in Fig. \ref{comp_exp_pm}.a, leading to significant distortion in the compressed signal, as shown later. In contrast, in the proposed compression, the source autoencoder learns how to reconstruct and compress the signal while avoiding the increase in its average power as explained in Section V.B. This is demonstrated in Fig. \ref{comp_exp_pm}.a, which shows that the majority of the amplitudes of the proposed-compressed signals are compressed and maintained below the saturation level.
\begin{figure*}[!htb]
\centerline{\includegraphics[width=1\textwidth]{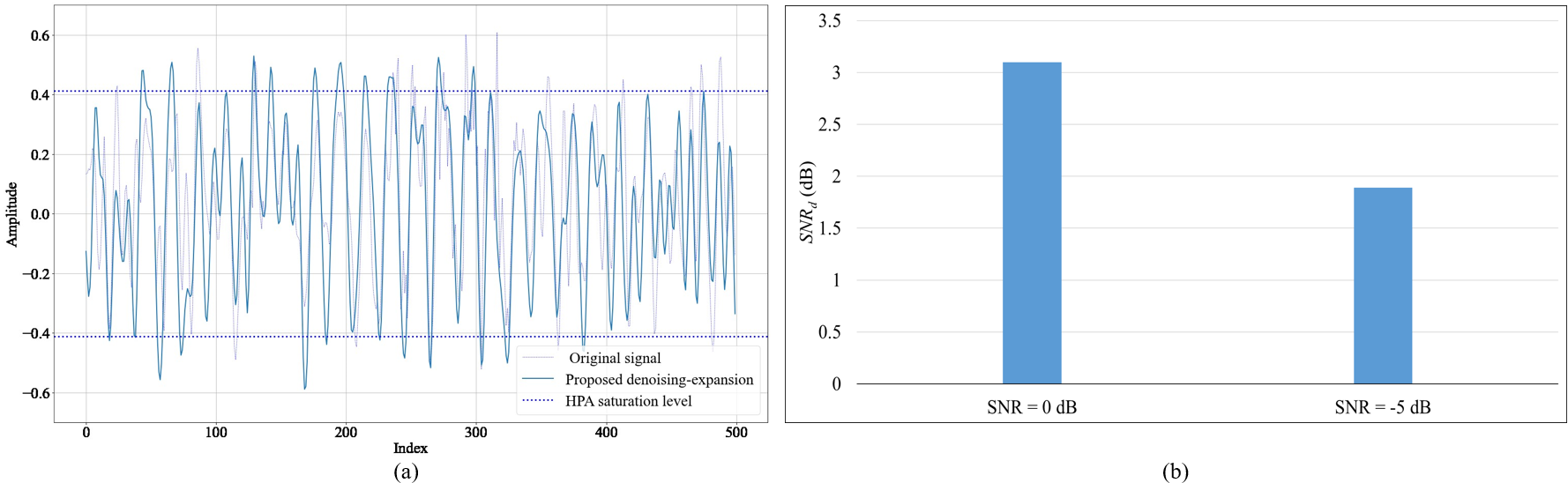}}
\caption{(a) a Denoised-expanded signal and its reference original signal. (b) Average $SNR_{d}$ of denoised-expanded signals.}
\label{denoise}
\end{figure*}

\subsubsection{\textbf{Average power of compressed signals}} Fig. \ref{pwr} depicts ratios of the mean average power of the compressed test signals with respect to the normalized average power of the original signals. As shown, the $\mu$-law compressed form of the vibration signal, on average, exhibits more than a 13-fold increment in its average power due to the enlargement of its small amplitudes. While this would reduce the quantization noise in analog-to-digital conversion, it would cause severe nonlinear distortion in the signal when passing through the HPA. Regarding the proposed compression, it slightly reduces the average power of the compressed form of the vibration signal. In terms of HPA nonlinearity, the slight decrease in average power shifts the input power of the compressed signals towards the linear region of the HPA. This is demonstrated in Fig. \ref{sp_hpa}, where the amplitude constellations of both the original and compressed test signals are displayed alongside the HPA's response curve. The HPA saturation level is also indicated as a dashed line on both axes. This visual setup provides useful insights into PAPR characteristics of the generated vibration signals, the behavior of the HPA, and the design requirements of the signal companding. Specifically, considering the peak amplitudes in the plot, it is obvious that all uncompressed signals and their $\mu$-law compressed forms will experience peak distortion after passing through the HPA since their peaks exceed the saturation level of the HPA. As for the mean amplitudes, all of the $\mu$-law compressed forms will experience significant and frequent amplitude distortion after passing through the HPA since their mean amplitude exceeds the saturation level. This is also the case for a considerable part of the uncompressed signals. On the contrary, the proposed framework compressed all the test signals so that their peaks and mean amplitudes fall below the saturation level. As a result, the compressed forms of the vibration signals are not subject to nonlinear distortion of the HPA. However, a slight amplitude distortion is still expected due to the soft limiting nature of the HPA and the imperfect reconstruction of the autoencoders. It is worth mentioning that the visual setup of Fig. \ref{sp_hpa} can be adapted to various VBCM systems to gain more insights into the PAPR characteristics, nonlinear behavior, and requirements for PAPR reduction.
\subsubsection{\textbf{Distortion due to nonlinear power amplification}}
Fig. \ref{exp} shows an uncompanded vibration signal before and after the HPA, its $\mu$-law companded form (\textit{the word companded is used here to refer to the signal that is compressed, passed through the HPA at the source, and expanded at the destination}), and its proposed-companded form. As shown, the uncompressed signal and its $\mu$-law companded form experienced a significant nonlinear distortion as their peak values are restricted to the HPA saturation level. While on the other hand, the amplitudes of its proposed-companded form are free of nonlinear distortion. Since the proposed framework compresses the signal so that its amplitudes fall in the linear region of the HPA, all amplitudes--- even the ones that exceed the HPA saturation level--- are restored after the expansion. The amplitude consultations of the uncompanded test signals, their $\mu$-law companded forms, and their proposed-companded forms are displayed in Fig. \ref{sp_pm_all}. By co-locating these constellations with the reference constellation as shown in Fig. \ref{sp_pm_all}.d, a convenient visual comparison to asses the nonlinear distortion can be made. The comparison clearly shows that the uncompanded signals and their $\mu$-law companded forms experience severe nonlinear distortion at the destination while, on the other hand, the proposed companding framework avoids nonlinear distortion and successfully restores the reference constellation---to a large extent--- at the destination. A comparison among the obtained EVM values is shown in Fig. \ref{evm}. The comparison demonstrates, in a quantified manner, the effectiveness of the proposed framework in mitigating the effects of nonlinear distortion. Specifically, while the uncompanded vibration signals and their $\mu$-law companded forms suffered from a very high distortion ($> 60\%$ EVM), the proposed-companded forms experienced very low distortion ($< 7\%$ EVM). As previously stated, the EVM quantifies the total system degradation experienced by the signals. For the uncompanded vibration signals and their $\mu$-law companded forms, the exhibited distortion is exclusively caused by the nonlinear power amplification. Regarding the prospered framework, the factors that contributed to the total system degradation are:
\begin{itemize}
\item  Soft limiting nature of the HPA: with $p=2$, the used SSPA model acts as a soft limiter.
\item  Autoencoder error: due to imperfect signal reconstruction of the autoencoders during compression/expansion stages. This error can be reduced by conducting more fine-tuning for hyperparameters of the autoencoders.  
\item  Noise presence and channel effects: While these impairments are not considered in the evaluation setup related to the obtained EVM results, they have a strong influence on the total system degradation in practical situations.
\end{itemize}
The obtained results from the EVM evaluation show that in the presence of nonlinear devices and the absence of a proper mechanism to reduce the PAPR, VBCM systems could suffer from severe nonlinear distortion. The results also confirm the effectiveness of the proposed framework in mitigating the effects of such distortion.

\subsubsection{\textbf{Spectral spreading}} Fig. \ref{psd} shows the mean PSD plots of the test vibration signals, their $\mu$-law-compressed forms, and their proposed-companded forms. The mean PSD of each of these three sets is calculated by estimating the individual PSD of each signal in the set after passing through the HPA. Accordingly, the mean PSD is obtained by averaging the estimated PSDs. The $\mu$-law-compressed forms experienced higher spectral broadening than the proposed-compressed forms. This regrowth in the spectrum is attributed mainly to the nonlinear distortion caused by the HPA. However, it should be mentioned that the logarithmic-based nature of the compression mechanism leads to spectral regrowth in the spectrum of the compressed signal.

\subsubsection{\textbf{Signal denoising}} To evaluate the denoising capability of the proposed framework, the test vibration signals are first compressed using the proposed compression, passed through the HPA, and then corrupted with the AWGN. Consequentially, at the destination, the compressed-amplified-noisy forms of the test signals are passed throughout the trained denoising-expanding autoencoder (trained according to the \textit{Noisy scenario} in Section.VI.B) to recover the original test signals. The plot in Fig. \ref{denoise}.a shows a recovered (denoised and expanded) signal alongside the reference original signal. By comparing both signals in the plot, it can be seen that the proposed framework is largely successful in removing the corrupted noise, expanding the compressed form, and restoring the original signal. The average $SNR_d$ values of the recovered signals are shown in Fig. \ref{denoise}.b. These values quantify the improvement in the recovered signals' SNR levels and confirm the proposed framework's capability in removing noise and mitigating the effects of noise expansion. Specifically, considering noisy, compressed signals with $0$ db and $-5$ db SNR levels, the improvement in the SNR after denoising and expending these signals is $3.1$ dB and $6.9$ dB, respectively. 

\section{Conclusion}
This paper has focused on addressing the PAPR of vibration signals and investigating the effect of nonlinear power amplification on the performance of VBCM systems. To the authors' knowledge, this study represents the first attempt in the context of VBCM to tackle the PAPR of vibration signals, explore the impact of nonlinear power amplification on the system, and propose a method for reducing the PAPR to mitigate the effects of nonlinear power amplification and enhance power efficiency. Specifically, the study conducted statistical analysis on the amplitude distribution of vibration signals and presented a closed-form formula to model the CCDF of the PAPR. Accordingly, analytical analysis and empirical investigation of the PAPR were conducted using various vibration datasets, which confirmed the occurrence of high PAPR values in vibration signals, particularly between $10$ to $13$ dB, when the number of samples exceeded $500$ in the acquired vibration segments. Moreover, the study examined the impact of nonlinear power amplification on system performance by utilizing signal constellation diagrams and error vectors as metrics to assess and quantify the degradation in the system. The findings revealed that HPA nonlinearity induces severe nonlinear distortion, resulting in reduced power efficiency. Consequently, the paper proposed a signal-companding-based framework for reducing the PAPR of vibration signals to mitigate the effects of HPA nonlinearity. This framework employed a lightweight reconstruction autoencoder with a compression-based activation function that simultaneously smooths and compresses the vibration signal while avoiding an increase in the average power of the compressed signal. Consequently, in the destination, the proposed framework used an expansion autoencoder that acts as a denoising filter to denoise the compressed signals before the expansion operation, thereby preventing the enhancement of the accumulated noise during signal expansion. The results demonstrated that the proposed framework significantly improved system performance and power consumption. In conclusion, this paper underscores the importance of implementing an appropriate signal companding mechanism in VBCM systems to mitigate the effects of nonlinear devices, ensuring efficient power consumption and reliable monitoring processes. \par

Future research directions would leverage Tiny Machine Learning (TinyML) techniques to facilitate the implementation of solutions that enable the proposed lightweight reconstruction autoencoder to perform inference tasks on the resource-constrained sensor nodes effectively. This will ensure the seamless execution of the autoencoder's operations in the remote sensor nodes, ultimately enhancing the applicability and practicality of the proposed framework.

\section*{Acknowledgment}

This work was funded in part by the National Research Council Canada under Project no.: AM-105-1.

\ifCLASSOPTIONcaptionsoff
  \newpage
\fi




\end{document}